\documentclass[10pt,preprint2,flushrt]{aastex}
\usepackage{epsfig}

\newcommand{\rr}{\mbox{\bf r}}
\newcommand{\HI}{\mbox{H\ {\sc i}}}
\newcommand{\N}{\mbox{\it N}}
\newcommand{\gadget}{\mbox{\sc gadget}}

\begin{document}
 
\title{Dynamical Evolution of Galaxies in Clusters}
 
\author{Oleg Y. Gnedin}

\affil{Princeton University Observatory, Princeton, NJ~08544\altaffilmark{1}}

\altaffiltext{1}{Present address: Space Telescope Science Institute,
                                  3700 San Martin Drive, Baltimore, MD 21218;
                                  ognedin@stsci.edu}

\begin{abstract}

Tidal forces acting on galaxies in clusters lead to a strong dynamical
evolution.  In order to quantify the amount of evolution, I run
self-consistent \N-body simulations of disk galaxies for a variety of
models in the hierarchically forming clusters.  The tidal field along
the galactic orbits is extracted from the simulations of cluster
formation in the $\Omega_0=1$; $\Omega_0=0.4$; and $\Omega_0=0.4$,
$\Omega_\Lambda=0.6$ cosmological scenarios.  For large spiral
galaxies with the rotation speed of 250 km s$^{-1}$, tidal
interactions truncate massive dark matter halos at $30 \pm 6$ kpc, and
thicken stellar disks by a factor 2 to 3, increasing Toomre's
parameter to $Q \gtrsim 2$ and halting star formation.  Low density
galaxies, such as the dwarf spheroidals with the circular velocity of
20 km s$^{-1}$ and the extended low surface brightness galaxies with
the scale length of $10-15$ kpc, are completely disrupted by tidal shocks.
Their debris contribute to the diffuse intracluster light.  The tidal
effects are significant not only in the core but throughout the
cluster and can be parametrized by the critical tidal density.  The
tidally-induced evolution results in the transformation of the
infalling spirals into S0 galaxies and in the depletion of the LSB
population.  In the low $\Omega_0$ cosmological models, clusters form
earlier and produce stronger evolution of galaxies.

\end{abstract}


\section{Introduction}

Hierarchical formation of clusters of galaxies involves strong local
variations of the potential field.  These tidal perturbations affect
the dynamics of infalling disk galaxies in a number of ways.  Firstly,
they lead to the truncation of stellar and dark matter distributions
at a critical radius of the Roche lobe.  Secondly, they increase the
stellar energy and the amount of random motion in the disk.  Tidal
heating may be strong enough to completely unbind small galaxies and
significantly transform larger ones.  Also, tidal heating is likely to
stabilize stellar disks against gravitational instability and suppress
subsequent star formation.  The result of all these effects may be the
morphological transformation of a spiral galaxy to an S0 type.

Such transformation is indeed revealed by the photometric and
spectroscopic studies of clusters of galaxies at various redshifts
(\citealt{JSC:00}, and references therein).  While the fraction of
elliptical galaxies stays roughly the same, the abundance of spirals
decreases by a factor of 2 to 3 from redshift 0.5 to the present.
Correspondingly, the number of S0 galaxies grows by the same amount.

In \cite{PaperI} I have argued that an important role in such a
transformation is played by the tidal interactions of galaxies with
their massive neighbors and with the global cluster halo.  Using
numerical simulations of the representative clusters of galaxies in
three different cosmological scenarios, I have estimated the amount of
tidal heating of the infalling galaxies.  The main contribution comes
from the interaction with the structures of order 100 kpc in size.

Previous studies (\citealt{R:76,FS:81}; and especially, Moore et
al. 1998, 1999; see more references in Paper I) have shown that close
encounters between disk galaxies can significantly modify their
structure and leave behind the distorted spheroidals.  The interaction
of galaxies with the cluster halo has also been studied \citep{M:84}
in the case of static spherical cluster models.  This leads to the
tidal truncation of galaxies.

In this paper, I investigate the dynamical effects of tidal heating of
disk galaxies in the hierarchically forming clusters.  I use a
high-resolution self-consistent \N-body code to resimulate a sample of
galaxies from the lower resolution cosmological simulations described
in Paper I.  The changes in stellar orbits result from the effects of
external tidal fields, computed from the second spatial derivative of
the cluster potential along the galaxy trajectories.  The fast variation
of the tidal force produces tidal shocks responsible for the
truncation of galactic halos and the kinematic heating of stars.  In
agreement with the estimates of Paper I, I find stronger tidal heating
in the low $\Omega_0$ cosmological scenario.

I use analytical arguments and discuss the expected evolution of
galaxies via tidal heating in \S\ref{sec:anal}.  In
\S\ref{sec:sample}, I describe the selection of the sample of galaxies
from the cluster simulations.  In \S\ref{sec:method}, I present the
numerical method and the tests of numerical relaxation.  Then I
describe the evolution of large spiral galaxies in \S\ref{sec:spiral},
including the possible effects on the gas.  The disruption of dwarf
and low surface brightness galaxies in \S\ref{sec:lsb} leads to the
discussion of the origin of intracluster light.  I compare these
results with the previous studies in \S\ref{sec:moore} and discuss
their observational implications in \S\ref{sec:dis}.

\section{The Critical Tidal Density}
  \label{sec:anal}

Let us start with an analytic estimate to indicate the range of
systems where tidal forces are important.  The external tidal force is
$-(d^2\Phi/dr_\alpha dr_\beta)_0 \, r_\beta \equiv F_{\alpha\beta} \,
r_\beta$.  The amplitude of the tidal field relates directly to the
tidal stripping density, which is analogous to the instantaneous
density of the Roche lobe around galaxies but depends on the highest
tidal peaks along the galactic orbit.  If a part of the galaxy has
lower density than the critical value, it would be stripped or torn
apart.  From Poisson's equation, the trace of the tidal tensor is
$F_{\rm tid} \equiv F_{\alpha\alpha} = 4\pi G \rho_{\rm tid}$.  A
typical value of the tidal peaks in Paper I is $F_{\rm tid} \sim 100$
Gyr$^{-2}$, and the corresponding tidal density is
\begin{equation}
\rho_{\rm tid} = 1.8 \times 10^{-3} \, \left({F_{\rm tid} \over
   100\ \mbox{Gyr}^{-2}}\right) \; M_{\sun}\, \mbox{pc}^{-3}.
  \label{eq:rhotid}
\end{equation}
This should be compared with the combined average density $\rho_{\rm
av} \equiv M(r)/(4\pi r^3/3)$ of stars and dark matter in a galaxy.

Consider a Milky Way-type galaxy with a large isothermal halo and an
exponential disk (as in \S\ref{sec:models}).  The average halo density
at the core radius is $\rho_{h,\rm av}(R_c) \approx 0.74 \,
\sigma_{250}^2 \, R_{c,\rm kpc}^{-2}\ M_{\sun}$ pc$^{-3}$, where
$\sigma_{250}$ is the velocity dispersion in units of 250 km s$^{-1}$.
The average disk density at one scale length is $\rho_{d,\rm av}(R_0)
\approx 0.63 \, M_{d,10} \, R_{0,\rm kpc}^{-3}\ M_{\sun}$ pc$^{-3}$,
where $M_{d,10}$ is the disk mass in units of $10^{10}\, M_{\sun}$.
Thus the inner regions of large spiral galaxies are stable against
tidal disruption.

Extended halos, falling outside the Roche lobe, would be truncated at
a radius $R_t$ such that $\rho_{\rm av}(R_t) \approx \rho_{\rm tid}$.
For our models this corresponds to the tidal radius $R_t \approx 42\,
\sigma_{250}$ kpc.  In addition to the instantaneous stripping,
secular tidal heating would enhance the mass loss.  Stellar disks with
$R_0 \sim 3$ kpc are better protected -- 99\% of their mass is within
the tidal radius.

On other hand, low surface brightness galaxies (including the dwarf
spheroidals) are in danger of being disrupted by tidal forces.  The
LSB galaxies have lower density than normal spirals, $\lesssim
10^{-2}\ M_{\sun}$ pc$^{-3}$, and are dominated by dark matter at all
radii.  Thus their disks would hold as long as they are being
protected by the halos.  Even though the halo velocity dispersions are
similar to those in large spiral galaxies, the larger sizes imply
lower density and so stronger truncation.  Stirred further by tidal
shocks, the LSB galaxies would lose most of their mass, leaving behind
extended tidal streams of stars and dark matter.  These qualitative
arguments are supported by detailed simulations in \S\ref{sec:spiral}
and \S\ref{sec:lsb}.

\subsection{Stripping of Dark Matter Halos}

The total amount of dark matter stripped from the galaxies can be
estimated analytically in the spirit of \cite{M:84}.  Assume that all
mass of the cluster belonged initially to the galaxies, distributed
according to the Schechter luminosity function
\begin{equation}
N(L) \, dL = N_0 \left({L \over L_*}\right)^{-\alpha} \, e^{-L/L_*}
     \, {dL \over L_*},
  \label{eq:lumfun}
\end{equation}
with $\alpha \approx 1$.  Let the mass-to-light ratio be constant and
use the \cite{FJ:76} relation for the velocity dispersion, $\sigma_g =
\sigma_{g,*} \, (L/L_*)^{1/4}$.  The characteristic break of the
luminosity function, $L_*$, corresponds to the mass scale $M_* \approx
5\times 10^{12} h^{-1} M_{\sun}$.  Finally, assume that the density
profiles of the galaxies and the cluster are isothermal with the
three-dimensional velocity dispersions $\sigma_g$ and $\sigma_{cl}$,
respectively.  The characteristic values in the simulations of Paper I
are $\sigma_g = 250$ km s$^{-1}$ and $\sigma_{cl} = \sqrt{3} \times
660$ km s$^{-1}$.  Then the galactic halos will be truncated at
\begin{equation}
R_t \approx {\sigma_g \over \sigma_{cl}} \, R_p,
\end{equation}
where $R_p$ is the distance of closest approach to the cluster center
(pericenter).  For many galaxies this distance is of the order the
cluster core radius, $R_p \sim 200$ kpc, which gives the truncation
radius $R_t \approx 44$ kpc.  The halo mass is then reduced to
\begin{equation}
M_t \approx {1\over G} \, {\sigma_g^3 \over \sigma_{cl}} \, R_p
    \approx 6.3 \times 10^{11}\ M_{\sun}.
\end{equation}
Tidal truncation modifies the mass function of galaxies such that the
effective exponential cutoff occurs at $M_t$ instead of $M_*$.
\cite{M:84} estimates the fraction of mass remaining bound to the
galaxies as 0.7 $M_t/M_*$, or about 6\% in our case.  Tidal heating
would further reduce the mass of the galaxies.

For a massive cluster CL 0024+1654, \cite{Tyson:98} were able to infer
the mass distribution using image reconstruction of the strong
gravitational lensing.  The potential of the cluster appears to be
very smooth, with only 2\% of the mass bound to luminous galaxies.
This is consistent with our estimate, as the velocity dispersion of CL
0024+1654 is twice that of our clusters.  Thus even if the galaxies
contained all of the cluster mass initially, they would retain only a
small fraction of it after tidal stripping.

\section{Constructing the Sample of Galaxies from the 
         Simulations of Hierarchical Cluster Formation}
  \label{sec:sample}

The sample of galaxies to be re-simulated is chosen from the lower
resolution cosmological simulations reported in Paper I.  The
Particle-Mesh (PM) simulations use constrained initial conditions to
produce representative clusters of galaxies for the three cosmological
scenarios: $\Omega_0=1$ (cluster I), $\Omega_0=0.4$ (cluster II), and
$\Omega_0=0.4, \Omega_\Lambda=0.6$ (cluster III).  All clusters have
the same one-dimensional velocity dispersion, 660 km s$^{-1}$, and
mass, $4\times 10^{14}\, M_{\sun}$.  The virial radii of the clusters
are \{1.0, 1.5, 1.4\} $h^{-1}$ Mpc, respectively.  The Hubble constant
is taken to be $H_0 = 65$ km s$^{-1}$ Mpc$^{-1}$ ($h = 0.65$).  The
mass function of identified halos follows the prediction of the
Press-Schechter theory for halos more massive than $3\times 10^{11}\,
M_{\sun}$ in Cluster I and $\sim 10^{12}\, M_{\sun}$ in Clusters II
and III.  Some of the lower mass halos are missing due to overmerging,
because the PM resolution is only 60 $h^{-1}$ kpc.  The orbits of 100
most massive galaxies in each cluster are traced through the
simulation starting from redshift $z=5$ using a subset of the most
bound galactic particles to calculate its center-of-mass motion.

The galactic trajectories calculated in Paper I are unique among other
cosmological simulations in that (i) they are real trajectories of the
halos in the evolving cluster, and not equilibrium orbits in the final
cluster potential; (ii) the motion of each identified halo is traced
through the simulation at each time step with an unprecedented
temporal resolution, $\sim 10^7$ yr; (iii) the trajectories are
calculated only for 1 Hubble time, which limits the average number of
orbital periods to 2--4; and (iv) the trajectories are affected by the
triaxiality of the cluster halo, which can make initially radial
orbits gain angular momentum and avoid the central regions.

The orbital eccentricity of a galaxy is defined as $\varepsilon \equiv
(R_a - R_p)/(R_a + R_p)$, where $R_a$ and $R_p$ are the median apo-
and pericenter distances, respectively, of the galactic orbit between
$z=5$ and $z=0$.  The median eccentricities for the sample of galaxies
in the three clusters are $\varepsilon_{\rm med} = \{0.27, 0.53,
0.38\}$.  They are lower than in the high-resolution simulation of
\cite{GHetal:98}, who integrated the orbits in the
spherically-symmetric potential approximating the cluster at the
present.  The different method of calculating orbits may be the cause
of the difference in the results.

In addition to the smaller eccentricities, the physical (not comoving)
pericenter distances of galaxies in Cluster I are also larger.  For
the three cluster samples, $R_{p,\rm med} = \{560, 160, 290\}$ kpc.
The pericenter distances may be relatively large because the clusters
move throughout the simulation as they form hierarchically.  Since the
galaxies are first identified at $z=5$, the cluster center of mass
travels on average $5.2$ comoving Mpc.  As a result of cluster
evolution, the galaxies often pass at a significantly larger distance
from the center than they would have in a fixed cluster model.

The resimulated galaxies are chosen from each of the three cluster
samples.  The galaxies ranked \#4 by mass represent large spirals and
are denoted \{L1,L2,L3\} for the cluster models \{I,II,III\},
respectively.  The orbital eccentricities of these galaxies,
$\varepsilon = \{0.27, 0.65, 0.38\}$, are typical of their clusters.
Two additional spirals, L3a and L3b (ranked \#13 and \#73), are chosen
from the Cluster III sample in order to investigate the variance of
the results within a given cosmological model.  In comparison with
galaxy L3, they have lower eccentricities ($\varepsilon = 0.27$ and
0.09, respectively) and smaller pericenter distances ($R_p = 220$ and
160 kpc versus $R_p = 230$ kpc for L3).

Three low surface brightness galaxies \{LSB1, LSB2, LSB3\} are modeled
using the same orbit and external tidal field of galaxy L3 but with
different masses and sizes.  Also, one galaxy from the bottom of each
cluster sample represents the dwarf spheroidal galaxies, \{D1,D2,D3\}.
Halos D1 and D2 are ranked \#93 by mass, while halo D3 is ranked \#87.
They all have low eccentricities, $\varepsilon = \{0.06, 0.09,
0.24\}$, and orbit in the outskirts of their clusters.

The initial parameters of the selected models are listed in Table
\ref{tab:initgal}.  The large galaxies are fairly massive spirals,
with the peak of circular velocity at 250 km s$^{-1}$.  The fiducial
value of the halo mass, $M_h = 8\times 10^{11}\ M_{\sun}$, is not much
lower than the true masses of L1-L3 in the cluster simulations ($\sim
2\times 10^{12}\ M_{\sun}$).  The fiducial dwarf model had to be
scaled down, because all identified halos in the cluster simulations
are relatively massive ($\gtrsim 10^{11}\ M_{\sun}$).  This would not
change the tidal histories of the dwarfs because they are still too
small to affect the evolution of the cluster potential.

\section{Numerical Method}
  \label{sec:method}

The dynamical evolution of galaxies is modeled using the
Self-Consistent Field method \citep{HO:92}.  In this type of
collisionless \N-body code, the potential and density are expanded in
a series of biorthogonal basis functions
\begin{eqnarray}
\Phi(\rr) & = & \sum A_{nlm} \, \Phi_{nlm}(\rr), \label{eq:potser} \\
\rho(\rr) & = & \sum A_{nlm} \, \rho_{nlm}(\rr),
\end{eqnarray}
which satisfy individually Poisson's equation:
\begin{equation}
\nabla^2 \Phi_{nlm}(\rr) = 4\pi G \rho_{nlm}(\rr).
\end{equation}
This assures that the linear combinations of the basis functions
satisfy Poisson's equation automatically.  The basis functions
employed by \cite{HO:92} consist of the ultraspherical polynomials of
$r$ and the spherical harmonics $Y_{lm}(\theta,\phi)$.  The expansion
is truncated at $n_{\rm max} = 10$ and $l_{\rm max} = 6$, with $m$
running as usual from $-l$ to $l$.

The SCF code has been modified for the current problem by \cite{GO:99}
and \cite{GHO:99}.  It has been optimized to run on parallel computers
and offers an almost ideal scaling with the number of CPU.  The
simulations were done on the SGI Origin 2000 supercomputer, using
$10^6$ particles each in the disk and in the halo.  The simulation
with $2\times 10^6$ particles takes about 20 hours on 16 R10000
processors.

The capability to use a large number of particles in the parallel SCF
code offers a high mass resolution of the galactic simulations.  The
masses of stellar particles in the models of large spirals are
$4\times 10^4\ M_{\sun}$ and in the models of dwarfs only $100\
M_{\sun}$.  Halo particles are more massive, however, in proportion to
the total halo mass.  Also, the simulations gain from the capability
to resolve temporal variations of the external tidal field on a scale
of $10^7$ yr.

All simulations start with a galaxy running in isolation for one
rotational period $T_{\rm rot}$ at the half-mass radius of the disk
($R_{1/2} \approx 1.7\, R_0$).  The particles reach dynamical
equilibrium but do not experience yet any appreciable heating.  Then
the external tidal field is applied for the amount of time elapsed
since $z=5$ until $z=0$ in the three cosmological scenarios ($T_{\rm
sim}$ in Table \ref{tab:fingal}).

The computations are done in dimensionless units, $G = M_d = R_0 = 1$.
Since the components of the tidal tensor $F_{\alpha\beta}$ have
dimensions of Gyr$^{-2}$, they are converted to the internal units
using only one parameter, $t_0 \equiv (G M_d/R_0^3)^{-1/2}$.  For
example, for the models of large spirals the conversion factor is $t_0
= 1.2 \times 10^7$ yr, or about 10\% of the rotational period.

\subsection{Initial Galactic Models}
  \label{sec:models}

The initial conditions for the disk galaxy simulations were generated
using a set of routines from \cite{H:93}.  The stellar disk is a
double-exponential, with the density profile
\begin{equation}
  \rho_d(R,z) = {M_d \over 4\pi R_0^2 \, z_0} \, e^{-R/R_0} \, e^{-|z|/z_0},
  \label{eq:rhod}
\end{equation}
where $M_d$ is the disk mass, $R_0$ is the radial scale length, and $z_0$
is the vertical scale height.  The dark halo is spherically symmetric and
corresponds to the isothermal sphere over some radial interval:
\begin{equation}
\rho_h(r) = {M_h \over 2\pi^{3/2}} \, {\alpha \over R_t \, (r^2 + R_c^2)}
   \, e^{-r^2/R_t^2},
\end{equation}
where $M_h$ is the halo mass, $R_c$ is the core radius, and $R_t$ is the
initial truncation radius.  The normalization constant $\alpha$ is
\begin{equation}
\alpha = \left( 1 - \sqrt{\pi} \, q \, e^{q^2} \,
                [1 - \mbox{\rm erf}(q)] \right)^{-1},
\end{equation}
where $q \equiv R_c/R_t$.  Halos are effectively truncated at $R_t$ to
reduce the computational task of following the orbits of very distant
particles.  In simulations with the external tidal field, the halos
are always stripped to a smaller radius so that the initial choice of
$R_t$ is unimportant.  For models L1--L3 and D1--D3 I take $R_t = 20
\, R_c$, and for models LSB1--LSB3 $R_t = 6 \, R_c$.  In all initial
models $R_c = R_0$.

\begin{figure}
\plotone{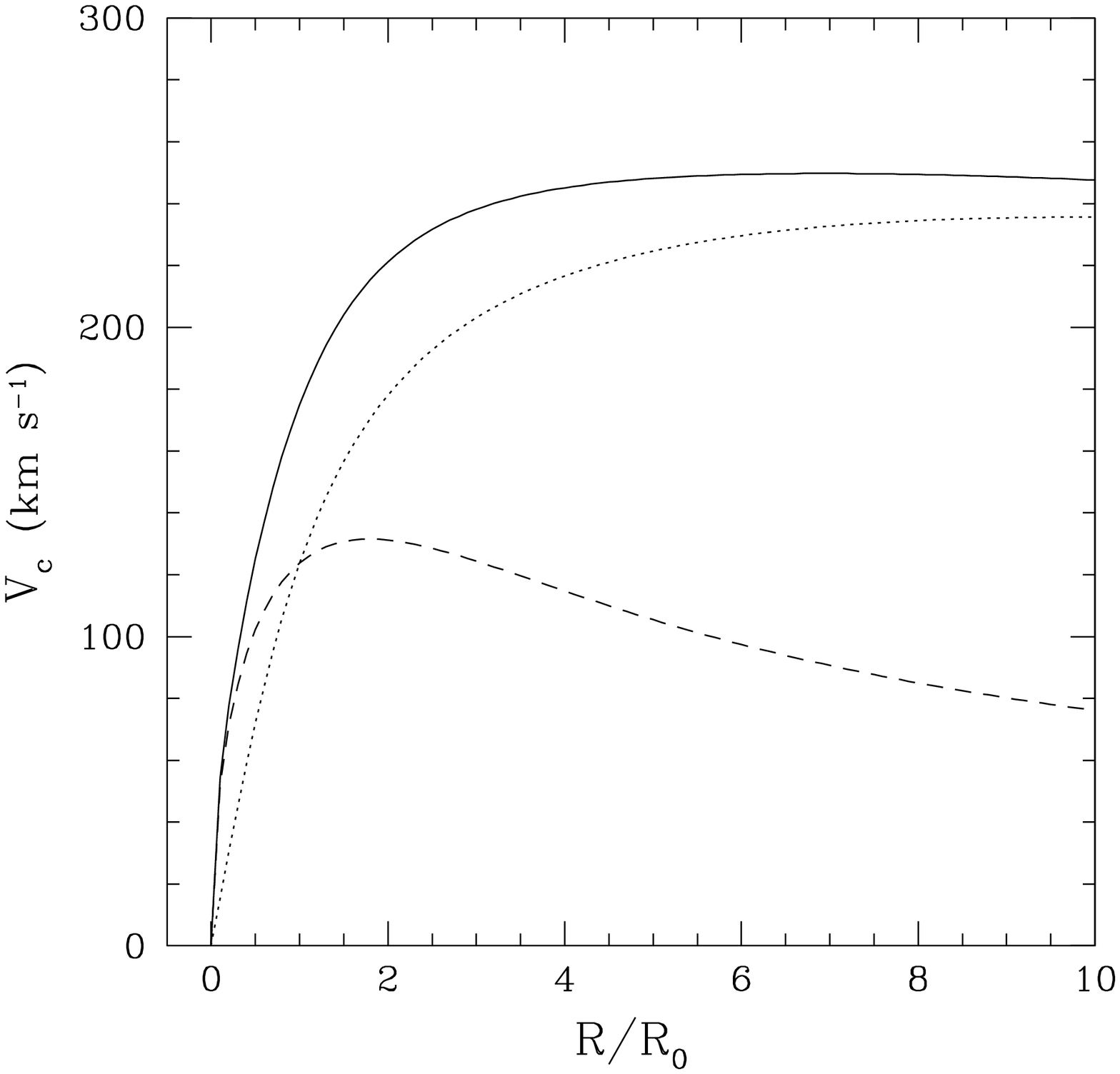}
\caption{Initial rotation curve of large disk galaxies (solid line) and the
  individual contributions from the exponential disk (dashes) and
  the isothermal halo (dots).  Cylindrical radius is in units of the disk
  scale length, $R_0$.
  \label{fig:vc_ini}}
\end{figure}

Figure \ref{fig:vc_ini} shows the disk and halo contributions to the
circular velocity
\begin{equation}
V_c(r) = \left( {G \, M(r) \over r} \right)^{1/2}
\end{equation}
of the initial models of large spirals.  The disk dominates over the
inner scale length, $R_0$, and the halo dominates at larger radii.
The combined rotation curve is flat out to $10 \, R_0$.

The figure of the halo is determined using the eigenvalues of the
reduced tensor of inertia, $I_{ij} \equiv \int x_i x_j dM$.  The
particle coordinates are rotated into a new frame where the inertia
tensor is diagonal.  The axis ratios of the triaxial ellipsoid are
calculated as $b/a = (I_{22}/I_{11})^{1/2}$ and $c/a =
(I_{33}/I_{11})^{1/2}$, with $c < b < a$.  The halos are initially
spherical with an isotropic velocity distribution.

The initial orientation of the galactic disk with respect to the
coordinate frame of the cluster simulation from Paper I can be set
arbitrarily.  I assume that the axes of the large galaxies are
parallel to the cluster frame and the axes of the dwarf galaxies are
tilted by $45^\circ$.  The tidal tensor has comparable components in
all directions, and therefore the results should not depend on the
initial disk orientation.

\subsection{Tests in Isolation}
  \label{sec:isolation}

Before studying the effects of tidal heating, it is very important to
assess the extent to which the simulations are affected by numerical
artifacts.  The main is numerical relaxation, which results in
erroneous heating of stellar orbits.  This relaxation arises from
Poisson fluctuations of the coarse-grained potential represented by a
finite number of particles, $N \sim 10^6$.  This is much smaller than
the number of stars in real galaxies, and therefore the relaxation
rate is artificially high (e.g. \citealt{W:96}).

I have run five test models in isolation in order to check the
dependence of numerical heating on the number of particles and the
time step (Tables \ref{tab:isola} and \ref{tab:isolb}).  The first
model, L--isol, is the main model for large spiral galaxies L1--L3 and
the next three are its modifications with reduced number of particles
or increased time step.  The last isolated model, LSB--isol3, is that
of a large low surface brightness galaxy LSB3.  The tests run for 12
Gyr, the longest simulation time of the three cosmological scenarios.

Due to the diffusive nature of the relaxation process, both the first
and second order changes of stellar energy are comparable, $\left<
\Delta E/E \right> \sim \left< (\Delta E/E)^2 \right>$.  However, the
former may change sign and vary randomly, whereas the latter rises
monotonically in time and offers a better measure of relaxation.  The
average energy change of disk particles at the end of run L--isol is
small, $\left< (\Delta E/E)^2 \right> \approx 0.5\%$.  Table
\ref{tab:isola} shows that the radial mass distribution of the disk
remains exponential and the scale length, $R_0$, changes only by 4\%.
The halo also remains spherical, with the axis ratios $b/a$ and $c/a$
close to unity.  The only noticeable difference is the vertical
heating of the disk.  The average scale height, $z_0$, increases by a
factor 1.7 and the vertical velocity dispersion $\sigma_z^2$ rises
similarly.  As Figure \ref{fig:heatzr} shows (lower dots), the
thickening is stronger near the center where the density of particles
is higher and the relaxation is expected to be faster.

Another statistic of disk heating is the change of the linear vertical
energy.  In equilibrium it can be defined using the epicycle theory as
\begin{equation}
  E_z = {v_z^2 \over 2} + {1\over 2}
        \left({\partial^2 \Phi \over \partial z^2}\right)_0 \, z^2,
  \label{eq:ez}
\end{equation}
where the second derivative of the potential is evaluated in the plane
of the disk (for a thin disk it is $2\pi G\Sigma(R)/z_0$, where
$\Sigma(R)$ is the surface density).  Equation (\ref{eq:ez}) is
strictly an integral of motion only near the midplane, where the disk
dominates gravity.  In the isolated runs the vertical energy $E_z$
changes significantly more than the total energy $E$, by the order
itself (Table \ref{tab:isolb}).  But this is consistent with the
changes of the scale height and the vertical velocity dispersion.

With half the number of particles in model L--isol2, the second order
energy change is 0.8\%.  The vertical energy and the scale height are
proportionately larger, while the radial scale length and the figure
of the halo are unchanged.  If only the number of disk particles is
reduced (model L--isol3), the heating is slightly less but similar to
the previous case.  Thus only the vertical structure of the disk is
sensitive to numerical relaxation, when the number of particles in
simulation $N \sim 10^6$.

The time step in the main simulations is $\Delta t = 0.1\, t_0$ in
code units, or about 1\% of the half-mass rotation period, $T_{\rm
rot} \equiv 2\pi R_{1/2}/V_c$.  In a thin disk the stellar orbits
should be calculated accurately even in the center, where the rotation
period decreases by 60\% as the disk density increases by a factor of
5.  However, once the relaxation raises velocities artificially high,
the stellar motion deviates more and more significantly from the
original orbits.  Doubling the time step in model L--isol1 leads to
larger energy changes than in the other isolated models.  This effect
is coupled to numerical relaxation and, therefore, both are more
important in the inner parts of the galaxy.  This is not a problem for
the present study because, by construction, external tidal forces
vanish in the center.  Tidal effects are more significant in the outer
regions of the galaxy where both numerical artifacts are less
important.

While the artificial thickening of the disk is certainly undesirable,
it can be corrected for in a statistical way.  The relaxation rate,
$d\left< (\Delta E/E)^2 \right>/dt$, is constant in time and the scale
height $z_{0,\rm num}$ increases linearly (cf. upper panel of Figure
\ref{fig:heatz}), as expected for pure Poisson fluctuations.  Thus at
any time it can be subtracted from the real simulation to accentuate
the effects of tidal heating.  Assuming that the numerical heating is
statistically independent from the external tidal heating, the two
effects add in quadrature and the modified scale height can be defined
as
\begin{equation}
  \bar{z_0} \equiv (z_0^2 - z_{0,num}^2 + z_{0,i}^2)^{1/2},
  \label{eq:z0num}
\end{equation}
where $z_{0,i}$ is the initial scale height.  The least squares fit to
the growth rate is
\begin{equation}
  d\ln{z_{0,num}}/dt = (6.0 \pm 0.05) \times 10^{-2}\ \mbox{Gyr}^{-1}.
  \label{eq:z0rate}
\end{equation}
In the following section we derive the rate of numerical relaxation in
a thin disk to confirm that the growth of $z_{0,\rm num}(t)$ is
consistent with the self-interaction of disk particles and with the
heating by more massive halo particles.

\onecolumn

\subsection{Numerical Relaxation in a Thin Disk}
  \label{sec:numrel}

\cite{SH:71} have derived the median relaxation time for a smooth
spherical system of mass $M$ and half-mass radius $R_h$:
\begin{equation}
  t_{rh} = 0.14 \, {N \over \ln{\Lambda}} 
           \left({R_h^3 \over G M}\right)^{1/2}.
  \label{eq:trh}
\end{equation}
It describes very well the rate of heating in simulations, even
including the resonant interactions of stars (see the discussion in
\citealt{GZ:02}, based on the results of \citealt{W:93}).  The
relaxation is dominated by distant encounters with large impact
parameters $b_{\rm max}$, parametrized by the Coulomb logarithm,
$\ln{\Lambda} = \ln{b_{\rm max}/b_{\rm min}}$.  The system relaxes on
many ($\sim 0.1 N$) dynamical times and for large enough $N$ can be
considered collisionless.

However, \cite{R:71} has emphasized that the relaxation time is much
shorter in the infinitely thin disks, of the order an orbital period
(see also \citealt{BT:87}, problem 8-6).  This is due to the close
encounters dominating in 2D geometry, instead of the distant
encounters in 3D geometry.  In a disk of finite thickness this effect
largely disappears, but there is another effect that leads to faster
relaxation.  The collective response of stars to a perturbation is
greatly enhanced when the stellar motions are highly ordered in a
kinematically cold disk \citep{S:87}.  Since the stellar orbits are
still largely confined to a plane, the most noticeable effect of disk
relaxation is the heating in vertical direction.

\subsubsection{Heating by Disk Particles}

For the general case of a thin disk with an initial scale height
$z_{0,i} \ll R_0$ and vertical velocity dispersion $\sigma_{z,i} \ll
V_c$, the relaxation rate can be estimated as follows.  The velocity
kick in the vertical direction from a single encounter with a star
$m_*$ at an impact parameter $b$ and relative velocity $v$ is $\Delta
v_z \approx 2 G m_* / v b$.  The encounter velocity is of the order
the radial and azimuthal dispersion in the plane.  Most of stellar
encounters take place at a median height above the plane, where the
density of the double-exponential disk (eq. [\ref{eq:rhod}]) is half
of that in the plane, $\rho_d(R,z_{1/2}) = 1/2 \times \Sigma(R)/2z_0$.
The integrated effect of encounters at all impact parameters is then
\begin{equation}
 {d\sigma_z^2 \over dt} = \int_{b_{\rm min}}^{b_{\rm max}}
   \left({2 G m_* \over v b}\right)^2 {\Sigma(R) \over 4z_0 m_*}
   v \, 2\pi b db
   = {2\pi G^2 m_* \Sigma(R) \ln{\Lambda_d} \over v z_0(t)},
  \label{eq:dsigmadt}
\end{equation}
where $b_{\rm max} \sim 2 z_0 \sim 0.6$ kpc and $b_{\rm min} \sim 0.1$
kpc (set by the resolution scale).  In the simplest case, as a result
of heating the surface density $\Sigma(R)$ remains the same but the
scale height $z_0$ increases in time.  The rate of increase can be
calculated from the relation $\sigma_z^2 = \pi G \Sigma(R) z_0$, which
should hold as long as the disk is still relatively thin:
\begin{equation}
 {d z_0 \over dt} = {1 \over \pi G \Sigma(R)} {d\sigma_z^2 \over dt}
   = {2 G m_* \ln{\Lambda_d} \over v z_0}.
\end{equation}
The solution is
\begin{equation}
 {z_0 \over z_{0,i}} = \left( 1 + {2 \, t \over t_{\rm rel,z}} \right)^{1/2},
\end{equation}
where the vertical relaxation time is
\begin{equation}
 t_{\rm rel,z} = {N_d \over 2 \ln{\Lambda_d}} {v \, z_{0,i}^2 \over G M_d}
   = 11 \, N_{d,6}\ \mbox{Gyr}
  \label{eq:trelz}
\end{equation}
for $v = 80$ km s$^{-1}$, $M_d \equiv N_d m_* = 4\times 10^{10}\,
M_{\sun}$, $z_{0,i} = 0.3$ kpc, $\ln{\Lambda_d} = 1.8$, and $N_{d,6}
\equiv N_d/10^6$.  The corresponding rise of the vertical velocity
dispersion is
\begin{equation}
 {\sigma_z \over \sigma_{z,i}} = \left({z_0 \over
   z_{0,i}}\right)^{1/2} = \left( 1 + {2 \, t \over t_{\rm rel,z}}
   \right)^{1/4}.
\end{equation}
This matches qualitatively the heating of the disk by massive
molecular clouds (cf. \citealt{L:84}).  The initial growth is linear
and the maximum rate is $d\ln{z_0}/dt = t_{\rm rel,z}^{-1} \approx
9\times 10^{-2}$ Gyr$^{-1}$, which is in reasonable agreement with
the observed rate of growth in the isolated simulation
(eq. [\ref{eq:z0rate}]).

If the stars were distributed spherically rather than in the disk, the
relaxation time would be significantly longer (eq. [\ref{eq:trh}]),
$t_{rh} \approx 10^{12}$ yr.  Alternatively, if the stars were
confined to an infinitely thin disk, the heating rate (in the plane)
would be
\begin{equation}
 {d\sigma_R^2 \over dt} = \int_{b_{\rm min}}^{b_{\rm max}}
   \left({2 G m_* \over v b}\right)^2 {\Sigma(R) \over m_*} v \, 2 db
   \approx {8 G^2 m_* \Sigma(R) \over v b_{\rm min}}.
\end{equation}
This rate (clearly dominated by close encounters) is faster than the
finite-thickness rate (dominated by distant encounters,
eq. [\ref{eq:dsigmadt}]) by a factor $2.1 \, z_0(t)/z_{0,i}$.  So as
expected, the finite-thickness disk provides an intermediate case
between the 2D and the 3D geometry.

\subsubsection{Heating by Halo Particles}

When the disk is surrounded by a dark halo, the heating by massive
halo particles can be more important.  The heating rate by the
particles with mass $m_h$ and density $\rho_h$ is
\begin{equation}
 {d\sigma_z^2 \over dt} = \int_{b_{\rm min}}^{b_{\rm max}}
   \left({2 G m_h \over v b}\right)^2 {\rho_h \over m_h}
   v \, 2\pi b db
   = {8\pi G^2 m_h \rho_h \ln{\Lambda_h} \over v},
  \label{eq:dsigmadt_h}
\end{equation}
where the encounter velocity $v$ is dominated by the halo dispersion,
$\sigma_h$.  The ratio of the heating rates due to the halo and disk
particles is
\begin{equation}
 {(d\sigma_z^2 / dt)_h \over (d\sigma_z^2 / dt)_d} =
    {m_h \, \rho_h \, \ln{\Lambda_h} \, v_{\rm enc,d} \over
     m_* \, \rho_d \, \ln{\Lambda_d} \, v_{\rm enc,h}}.
\end{equation}
The typical parameters in our models at $R_0$ are $m_h / m_* = 20$,
$v_{\rm enc,h} / v_{\rm enc,d} \approx 2$, and $\rho_d / \rho_h
\approx 5-10$, so the two heating rates are comparable.  However, the
growth of the scale height is different because now the right-hand
side of equation (\ref{eq:dsigmadt_h}) is independent of $\sigma_z$
and
\begin{equation}
 {z_0 \over z_{0,i}} =
   \left({\sigma_z \over \sigma_{z,i}}\right)^2 =
   1 + {t \over t_{\rm rel,h}}
\end{equation}
with
\begin{equation}
 t_{\rm rel,h} =
   {N_h \over 8\pi \ln{\Lambda_h}}
   {\sigma_{z,i}^2 \, v \over G^2 M_h \rho_h}.
  \label{eq:trelh}
\end{equation}
This interaction with the halo particles is equivalent to the heating
by massive black holes (cf. \citealt{LO:85}).  For $M_h = 8\times
10^{11}\, M_{\sun}$, $v \approx 160$ km s$^{-1}$, $\sigma_{z,i}
\approx 80$ km s$^{-1}$, $\ln{\Lambda_h} \approx 2$, and $\rho_h(R_c)
\approx 0.08\, M_{\sun}$ pc$^{-3}$, the relaxation time is $t_{\rm
rel,h} \approx 17 \, N_{h,6}$ Gyr.  This is in excellent agreement
with the growth rate in the isolated simulation
(eq. [\ref{eq:z0rate}]).  Thus, both disk and halo particles
contribute to the disk thickening, although in the long run the
heating by halo particles would dominate.

Having a certain understanding of the numerical relaxation in the
disk, when it sets in (eqs. [\ref{eq:trelz}] and [\ref{eq:trelh}]) and
how to account for it (eq. [\ref{eq:z0num}]), we move to the
simulation results.

\twocolumn

\section{The Transformation of Large Spiral Galaxies}
  \label{sec:spiral}

We start with the results for large disk galaxies.  Figure 2 shows the
distribution of particles at the beginning and the end of the
simulation of galaxy L1.  The first important result is that the
galaxy survives tidal heating.  It still looks like a disk galaxy of
roughly the same size.  Large, coherent spiral arms form in the outer
regions.  The disk shrinks in the radial direction and flares at the
fuzzy edges.  But the main difference from the initial configuration
is a substantial thickening of the disk.  Viewed at a certain surface
brightness limit, this thick and round disk galaxy may be classified
as an S0.

The dark matter distribution is relatively simple.  The galaxy has
been stripped of about 40\% of its halo (Table \ref{tab:fingal}) and
has shrunk to the limits of the stellar disk.  Inside the truncation
radius, the halo has been little affected (Figure \ref{fig:denh}).
The figure of the halo becomes triaxial but only mildly, with the
smallest axis ratio $c/a = 0.88$.  Thus, the amount of the remaining
halo mass is indicated simply by the truncation radius, $R_t$.  For
galaxy L1 it changes from 60 kpc to 37 kpc.  Figure \ref{fig:vcirc}
shows that the rotation curve is still flat up to six disk scale
lengths.  The circular velocity, measured at $5 R_0$, is decreased
only by 2\%.

The disk of the galaxy is perturbed more significantly.  Figure
\ref{fig:dend} shows that the surface density decreases in the center
but increases between $3 R_0$ and $4 R_0$, relative to the initial
profile.  The surface density exhibits noticeable wiggles superimposed
on the otherwise exponential disk.

In addition to truncating the halo, a major effect of tidal heating is
the thickening of the disk.  By construction, the external tidal force
vanishes at the center and increases linearly with radius.  Figure
\ref{fig:heatzr} demonstrates that the disk flares in the outer parts.
Numerical relaxation affects the disk structure only within the inner
two scale lengths, where tidal heating is negligible.  At $R > 2 R_0$,
the external tidal effects are clearly dominant.

A direct evidence of tidal heating is the dramatic increase of all
three velocity components in the outer regions of the disk (Figure
\ref{fig:dispd}).  The velocity dispersion increases at the expense of
the rotation speed, which starts to decline at around four scale
lengths from the center (while the circular velocity is still flat;
cf.  Figure \ref{fig:vcirc}).  In contrast, the velocity dispersion of
the halo decreases in the outer parts due to the significant mass
loss.

The analysis of the other large spirals shows a lot of similarity but
with quantitative differences.  Since the tidal field is stronger in
the low $\Omega_0$ clusters (see \S\ref{sec:tidhis}), galaxies L2 and
L3 are subject to stronger perturbations than galaxy L1.  Galaxy L2
suffers the biggest mass loss, about 64\%, while galaxy L3 provides
the intermediate case with about 50\%.  Their truncation radii fall
below 30 kpc.  The disks of both galaxies L2 and L3 are heated more
strongly than in galaxy L1.

Figure \ref{fig:halom} shows that the mass loss from the three halos
is not steady.  There are several instants of large losses associated
with the maxima of the tidal force.  Examining the particle
distribution at successive epochs reveals how the halo is being
stripped.  After a strong tidal shock, a cloud of dark matter
particles disperses across the tidal boundary and slowly drifts away
from the galaxy in two opposite directions along the same line.  That
line corresponds to the direction of the strongest component of the
tidal tensor at the time.  The unbound particles are later stretched
along that direction further and further away from the galaxy.

As Figure \ref{fig:heatr} shows, the disk scale lengths of all three
galaxies remain almost constant.  In contrast, the vertical scale
heights rise monotonically with time (Figure \ref{fig:heatz}).  While
a part of this effect is due to the numerical relaxation
(eq. [\ref{eq:z0num}]), the corrected scale height $\bar{z_0}$
measures the thickening only due to the tidal heating.  Galaxy L2
leads in the rate of expansion, in agreement with other indicators of
evolution, the mass loss and the velocity dispersion increase.  At the
end of the run, its scale height is more than doubled.  The other two
galaxies experience more modest evolution.  In particular, the disk of
galaxy L3 had expanded to its maximum height in 7 Gyr and then
remained effectively unchanged for the same amount of time until the
present.

\subsection{Dependence on Time Step and Number of Particles}

In order to quantify the dependence of the results on numerical
details, such as the time step and the number of particles, I have run
three additional models of galaxy L3 (Table \ref{tab:L3gal}).  They
are simulated using the same parameters as the isolated models
\{L--isol1, L--isol2, L--isol3\} but with the external tidal field of
galaxy L3.  The corresponding isolated runs are used to correct the
scale height $\bar{z_0}$.

In model 1, the time step is doubled while the number of particles is
kept the same, relative to the main model of L3.  As discussed in
\S\ref{sec:isolation}, the time step is chosen small enough to
represent accurately stellar orbits even in the center; $\Delta t =
0.01 \, T_{\rm rot} \approx 10^6$ yr.  Indeed, with the doubled time
step model 1 reproduces most of the features of model L3.  The main
changes are the 13\% stronger thickening of the disk and a slightly
longer radial scale length.

In model 2, half the number of particles is used in both the disk and
the halo with the original time step.  The results are very close to
those of model L3.  The changes of the scale length and the corrected
scale height are not noticeable at all.

In model 3, where only the number of disk particles is reduced by
half, the scale height grows even less than in model L3.  This
indicates that most of the numerical heating of disk particles is
produced by the interaction with massive halo particles, as expected.
In all three test runs the halo structure and mass loss are virtually
identical.

The conclusion from these tests is that only the vertical structure of
the disk is sensitive to numerical effects.  Variations of the time
step and the number of particles affect the corrected scale height
$\bar{z_0}$ by 10\% to 30\%.  Thus the deviation of our main results
from the case of infinite resolution is expected to be of the same
order.

\subsection{Dependence on Force Resolution}

The necessary truncation of the spherical harmonics series in equation
(\ref{eq:potser}) imposes a limit on the vertical force resolution in
a thin disk.  An accurate calculation of the force with $z_0/R_0 =
0.1$ may require the expansion order $l_{\rm max} \sim 100$.  Our
simulations with $l_{\rm max}=6$ thus have an effectively smoothed
vertical force within the disk, and the comparison with the forces
calculated by direct summation shows significant deviations at $\left|
z \right| \lesssim 0.4 \, R_0$.  From the SCF simulations alone it is
difficult to assess how this could affect the response of the disk to
external perturbations.  In order to test sensitivity of the results
to force resolution, I use the publicly-available tree code \gadget\
\citep{Gadget} and resimulate large galaxy models L1, L2, and L3.

The tree algorithm with an accuracy parameter $\theta=0.8$ calculates
the vertical forces in our thin disks within 1\% of the direct
summation result.  The algorithm is, however, much more
computationally expensive and can be realistically used only with a
subset of particles.  With $N_d = 10^5$, $N_h = 4 \times 10^5$, one
parallel \gadget\ simulation takes 150 hours on 8 processors of the
beowulf PC cluster at STScI.  I use a 'type 0' time step criterion
based on local particle acceleration $a$ and softening parameter
$\epsilon$: $\Delta t \propto \sqrt{\epsilon/a}$, and limit the time
step not to exceed 0.1 in code units ($\approx 10^6$ yr, the same step
as in the SCF runs).  The average time step was $0.04-0.08$ in code
units.  In order to adequately resolve the vertical structure of the
disk and to minimize relaxation effects, I take the softening
parameter $\epsilon_d = 0.1 z_0$ for disk particles and $\epsilon_h =
0.1 R_0$ for halo particles, using the definition of \citet{Gadget}.

Since the numerical heating of the disk is largely due to the massive
halo particles, I run two sets of models increasing $N_h$ from $10^5$
to $4\times 10^5$.  The number of disk particles is fixed at $N_d =
10^5$ (these numbers are limited by a realistic run time).  In the
control run in isolation with $N_h = 10^5$ the disk size shows a large
increase ($z_0/z_{0,i} = 3.0$) but so does a corresponding SCF run
with the same number of particles ($z_0/z_{0,i} = 3.1$).  In the
control run with $N_h = 4\times 10^5$ the numerical heating is already
reduced ($z_0/z_{0,i} = 1.9$) almost to the level of the main SCF
runs.  In both isolated \gadget\ simulations the radial scale length
changes by less than 1\%.  This agrees with the discussion in
\S\ref{sec:numrel} that numerical heating depends primarily on the
number and masses of particles.

Table \ref{tab:gadget} shows the results of \gadget\ simulations for
models L1--L3.  Overall, the disk heating is somewhat stronger than in
the SCF runs: both $\bar{z}_0$ and $R_0$ increase by roughly 10\%.
The vertical heating is much stronger in the runs with fewer halo
particles -- this is clear evidence that numerical relaxation can
affect the evolution of thin disks even with adequate force
resolution.

The thickening of the disk is less monotonic in the \gadget\ runs.
Sudden jumps of $\bar{z}_0$, for example at 5 Gyr and 9 Gyr for model
L2, coincide with larger than usual deviations of the disk
center-of-mass from the halo center-of-mass.  The two can differ by
$\Delta R_{CoM} \lesssim 800$ pc after a strong perturbation, and in
such an unrelaxed state the disk appears thicker.  Note that this
deviation is still within $0.3 \, R_0$.  In models L1 and L3, and in
all SCF runs, $\Delta R_{CoM} < 100-200$ pc.

The disk appears symmetric in the vertical direction and shows no sign
of warping.  The orientations of the inner and outer parts of the
disk, divided at $2.5 R_0$, differ by less than $5\degr$ ($8\degr$ for
SCF runs).

A noticeably difference in the \gadget\ runs is the radial expansion
of the disks.  In the SCF models, $R_0$ was essentially unchanged.
This might be a result of the restoring forces being effectively
``anisotropic'' within the disk: while the radial expansion is
sufficiently accurate, the vertical expansion is smoothed.
Fortunately, the difference with the \gadget\ results is only
$10-15\%$.

As a result of the increase of $R_0$, the circular velocity drops
correspondingly by about 10\% in all \gadget\ models.  However, the
halo structure in the \gadget\ runs almost exactly mirrors that in the
SCF runs.  The only deviation of the halo mass is in model L3 (by
3\%), while the shapes of the halos are similar.  I conclude that
apart from $R_0$, the only quantity susceptible to the errors in force
resolution is $\bar{z}_0$, and while it increases by a factor
$1.6-2.4$ in the SCF runs the true value is probably in the range
$2-3$.

\subsection{Dependence on Tidal History}
  \label{sec:tidhis}

In cluster simulations presented in Paper I, the amount of tidal
heating has been estimated using the tidal heating parameter, $I_{\rm
tid}$.  It is simply the square of the integrated velocity change
summed over all peaks of the tidal force $F_{\alpha\beta}$
($\alpha,\beta=\{x,y,z\}$):
\begin{equation}
I_{\rm tid} \equiv \sum_n \sum_{\alpha,\beta} \left( \int
  F_{\alpha\beta} \, dt \right)_n^2 \, 
  \left( 1 + {\tau_n^2 \over t_{\rm dyn}^2} \right)^{-3/2}.
  \label{eq:ipar}
\end{equation}
It is also corrected for the conservation of adiabatic invariants of
stellar orbits when the external perturbations are long, using the
parametrization of \cite{GO:99}.  Here $\tau_n$ is the effective
duration of peak $n$ for each value of $\alpha$ and $\beta$, and
$t_{\rm dyn}$ is the half-mass dynamical time of the galaxy ($\sim
T_{\rm rot}$).  The average energy change due to tidal heating for the
stars at radius $r$ is
\begin{equation}
\langle\Delta E\rangle = {1\over 6} \, I_{\rm tid} \, \langle r^2 \rangle.
\end{equation}
This semi-analytical estimate provides a good description of the
energy changes of particles in the simulation and can be used to
compare tidal effects for different galaxies.

The galaxies in each cluster sample show an exponential distribution
of the tidal parameter, $N(I_{\rm tid}) = N_0 \, \exp(-I_{\rm
tid}/I_0)$.  The characteristic parameter $I_0$ is the largest in
Cluster II ($5 \times 10^3$ Gyr$^{-2}$), intermediate in Cluster III
($2 \times 10^3$ Gyr$^{-2}$), and the smallest in Cluster I ($8 \times
10^2$ Gyr$^{-2}$; see Paper I).  The simulations of galaxies L1--L3
show noticeable differences in the evolution of large spiral galaxies
in the three cosmological models.  In agreement with the estimates of
the tidal heating parameter, galaxy L2 experiences stronger evolution
than galaxies L1 and L3.

However, these differences may not serve as a clear discriminant among
the cosmological models.  The distribution of tidal parameters within
the same cluster leads to a comparable variance of the results.  The
additional simulations of two galaxies from Cluster III prove this
point.

The two computed models, L3a and L3b, use the same initial conditions
as galaxy L3 but different tidal histories.  These galaxies,
corresponding to the Cluster III halos \#13 and \#73, have some of the
largest heating parameters, $I_{\rm tid} = 3.8\times 10^3$ Gyr$^{-2}$
and $I_{\rm tid} = 6.4\times 10^3$ Gyr$^{-2}$, respectively.  For
comparison, the reference galaxy L3 has $I_{\rm tid} = 4.2\times 10^3$
Gyr$^{-2}$.  These models represent the strongest cases of tidal
evolution.

As expected, galaxy L3b loses more mass and develops a thicker disk
than model L3, while for L3a the trend is the opposite (Table
\ref{tab:fingal}).  The final state of galaxy L3b is similar to that
of model L2 from Cluster II.  On the other hand, galaxy L3a is closer
to model L1 from Cluster I.  Thus the distribution of tidal histories
of galaxies within the same cluster leads to a similar variation of
the dynamical evolution.

The adiabatic correction factor in equation (\ref{eq:ipar}) is very
important.  It accounts for the temporal structure of the tidal field
and suppresses the perturbations with long timescales to which
galaxies respond adiabatically.  Without that factor, galaxy L3a would
have had a higher value of $I_{\rm tid}$ than galaxy L3b.  The
difference is due to the tidal force being more ``impulsive'' for
galaxy L3b, while being more ``adiabatic'' for L3a.

\subsection{Do spirals become S0 galaxies?}

The simulations of large spiral galaxies show that the kinematic
heating and vertical expansion of the disks lead to a significant
morphological transformation.  An initially normal spiral may be
identified in the end as a lenticular galaxy, an S0.  The following
section puts this claim to a thorough observational check.

The final snapshot of galaxy L1 (Figure 2) looks similar to the deep
exposures of the S0 galaxy NGC 4762 \citep{S:61,B:79}.  The disk is
thick and ends with fuzzy warps.  A modest central mass concentration
is seen in both the simulation and the real galaxy.

The most detailed data are available for nearby S0 galaxies in the
field \citep{Kent:85}.  The HST images of lenticulars in distant
clusters \citep{Eetal:97,vDetal:98} usually provide only integrated
light and colors.  Typically, the surface brightness profiles of S0s
are well fitted by a combination of the exponential disk and the
$r^{1/4}$ bulge.  The simulations reported here do not include a bulge
component initially.  In fact, one of the goals of this study was to
investigate whether a bulge forms as a result of tidal heating.  Two
competing effects play a role here: stars on average gain energy from
tidal perturbations and move generally outward; on the other hand, the
tidally-induced energy dispersion contributes to two-body relaxation
and accretion onto the center.  Figure \ref{fig:dend} shows that no
appreciable bulge forms from the disk particles, asserting that the
relaxation time is still much longer than the age of the galaxy.
However, the interstellar gas may be driven inward by tidal effects
and ignite a burst of star formation (\citealt{HM:95}; see also
\S\ref{sec:gas}).  This would assist in the formation of a prominent
bulge.

S0 galaxies typically have the same disk scale lengths as the spirals
of similar luminosity \citep{Kent:85}.  In the present simulations the
scale length $R_0$ is virtually unchanged.  Also, the axis ratios of
the simulated disks, 0.15 to 0.25, are consistent with the average
observed value of 0.2-0.25 for the ``thick disks'' \citep{BM:98}.
\cite{B:79} attempted to decompose the vertical brightness profile of
late type edge-on galaxies into a combination of the thin and thick
disks.  The latter dominate in the outer regions of the galaxies and
are much more prominent in S0s than in spirals.  Thick disks are not
as centrally peaked as thin disks and have a shallow radial gradient:
they look ``diffuse'' and ``boxy''.  All this is consistent with
having a tidal origin; flaring of the disk was expected and has been
detected in the simulations.

Kinematic information is available only for the local galaxies.  In a
cold exponential disk, vertical velocity dispersion should fall with
radius as the surface density: $\sigma_z^2(R) = 2\pi G z_0 \Sigma_0
\exp(-R/R_0)$.  In a sample of ten S0 galaxies \citep{SS:96} the
velocity dispersion flattens at $50-100$ km s$^{-1}$ and even rises in
the outer regions.  Similarly, a large study of the early type
galaxies by \cite{SP:98} shows a systematic increase of the total
velocity dispersion at large radii, often exceeding 100 km s$^{-1}$.
Present simulations certainly confirm this trend.

Finally, note that bars are commonly found in S0 galaxies.  A single
tidal interaction may excite bar instability in an axisymmetric
galaxy, but a succession of tidal shocks and the resulting kinematic
heating would isotropize stellar orbits and destroy the bar.  The
actual survival of bars, then, depends on the amount of tidal heating
in the inner galaxy (see \citealt{S:99}).  Present simulations, which
include a stabilizing dark matter halo, do not show any detectable
bar.

\subsection{Gas Dynamics and Star Formation}
  \label{sec:gas}

The evolution of spiral galaxies depends sensitively on the amount and
distribution of the interstellar gas.  Even though the collisionless
simulations do not address gas dynamics directly, it is possible to
infer qualitative features of the gas response to tidal heating.

The external tidal force is not steady in time, it comes in a series
of peaks (cf. Figures 14--16 in Paper I).  The
timescale of each tidal perturbation is of the order $10^8$ yr or
longer.  While the stars would continue to gain energy without
dissipation, the gas should cool to its equilibrium temperature fairly
quickly.  The cooling time of the ionized gas is
\begin{equation}
t_{\rm cool} = {\frac{3}{2} nkT \over n_e^2 \, \Lambda(T)}
	 \approx 7\times 10^5 \, {T_6\over n_e \, \Lambda_{-23}} \ \mbox{yr},
\end{equation}
where $\Lambda(T) \sim 10^{-23}$ erg cm$^3$ s$^{-1}$ is the cooling
function at the temperatures $T \sim 10^6$ K, and $n$ and $n_e$ are
the baryon and electron number densities, respectively.  As a case of
extremely fast tidal heating, consider that the gas with $n T \sim
10^4$ K cm$^{-3}$ heats up to $8\times 10^5$ K, corresponding to the
velocity dispersion of 100 km s$^{-1}$.  Even in this case, it would
cool back to $10^4$ K in $4\times 10^7$ yr.  In reality the gas would
likely heat less efficiently and therefore cool faster.

However, the thick stellar disk is now stable against gravitational
perturbations and is unlikely to form new stars.  Figure \ref{fig:q}
shows Toomre's $Q$ parameter \citep{T:64} for the final states of
galaxies L1--L3.  Even the minimum value, $Q_{\rm min} \approx 2$, is
large enough to suppress any axisymmetric disk instability.  Under
such conditions spiral structure and gaseous shocks would not form in
the inner regions of the galaxies.  The spiral arms visible in Figure
2 are prominent only in the outer regions, where the diffuse gas is
likely to be ionized or stripped by the intracluster medium.

The stability of stellar disks is modified in the presence of cold
gas.  The low velocity dispersion of the gas amplifies the growth of
perturbations and lowers the effective value of $Q$ compared to the
pure stellar case.  \cite{JS:84} derive a stability criterion for a
two-fluid system of the stars and gas.  From their equation (22) it
follows that the system is unstable if the modified parameter $Q_{\rm
eff}$ is less than unity:
\begin{equation}
Q_{\rm eff} = Q \, {3.36 \over 2\pi} \, \left({x \over x^2+1} +
    {\beta x \over x^2 + c}\right)^{-1} \, < 1,
  \label{eq:qeff}
\end{equation}
where $\beta = \Sigma_{\rm gas}/\Sigma_*$ is the ratio of the gas to
stellar surface densities, $c = v_{\rm th}/\sigma_*$ is the ratio of
the gas thermal velocity to the stellar velocity dispersion, and $x =
\kappa/(\sigma_* \, k)$ is the ratio of the perturbation wavelength,
$2\pi/k$, to the effective scale of stellar motion.  Here
\begin{equation}
\kappa \equiv
    \left({1\over R} {dV_c^2 \over dR} + 2 {V_c^2 \over R^2}\right)^{1/2}
\end{equation}
is the epicyclic frequency, which includes the contributions from both
stars and dark matter.  Equation (\ref{eq:qeff}) allows one to
determine the amount of gas, $\beta$, required to make the disk
unstable.  Writing the velocity dispersions as $v_{\rm th} = 11\
T_4^{1/2}$ km s$^{-1}$ and $\sigma_* = 100\ \sigma_{*,100}$ km
s$^{-1}$, I find $c = 0.11\ T_4^{1/2} \, \sigma_{*,100}^{-1}$.  The
most unstable mode is usually a half of the maximum allowed wavelength
\citep{BT:87}, which in this case is $x_{\rm max} \approx 1.87
(1+\beta)/Q$.  With these parameters, the equation $Q_{\rm eff}=1$ can
be solved numerically to yield $\beta \approx 0.52$.  It shows that
the disk is unstable only when the mass of cold gas is at least 50\%
of the stellar mass, a condition rarely satisfied in any spiral or S0
galaxy even in the field.

Current theories of star formation in normal spiral galaxies often
rely on a continuous infall of cold gas (e.g.,
\citealt{Eetal:93,DSS:97}).  This infall supplies the material for new
stars and also helps to sustain a marginal dynamical instability of
the disk.  After galaxies enter the cluster, external tidal forces
halt the late infall and, with it, new star formation.

\subsection{Ram Pressure Stripping}

The amount of gas remaining in the galaxies can also be reduced by a
variety of processes.  The intracluster media (ICM) can sweep the
diffuse gas via ram pressure \citep{GG:72}.  Assuming the ICM density
in the cluster core $n_{\rm icm} = 10^{-3}$ cm$^{-3}$ \citep{JF:84},
the surface density of \HI\ gas $\Sigma_{\rm gas} = 4\ M_{\sun}$
pc$^{-2}$ \citep{BM:98}, and the orbital velocity of 800 km s$^{-1}$,
large spirals with exponential disks would be stripped of atomic gas
down to $R = 5$ kpc.  Recent numerical simulations of ram pressure
stripping in rich clusters \citep{AMB:99} confirm that the \HI\ disks
are truncated at $\sim 4$ kpc.  Through viscous coupling even the
inner gas may be affected \citep{QMB:00}.

The efficiency of ram pressure stripping depends also on the
hierarchical formation of the cluster.  When individual groups of
galaxies merge into the cluster, a bow shock may form slightly ahead
of the group's motion.  It happens when the ram pressure on the
infalling gas increases faster than a sound wave can propagate across
the substructure \citep{FD:91}.  Numerical simulations \citep{RLB:97}
confirm that bow shocks do form around the substructures falling in at
supersonic speeds.  These shocks may delay the heating of gas to the
virial temperature.  X-ray observations of the central region of Coma
cluster by {\it ASCA} \citep{HDetal:99} find a cold and a hot spots in
the temperature map.  A natural explanation would be a bow shock
associated with the group around the bright galaxy NGC 4869.

While ram pressure stripping does not affect directly dense molecular
clouds, the interpenetrating galactic encounters may remove the
molecular component \citep{K:83}.  Direct collisions between molecular
clouds would be rare but the coupling via magnetic fields would ensure
dynamical interaction.  In high-speed collisions the gas would shock
and decelerate, whereas the stars and dark matter fly away.

In a survey of 17 bright spirals in the Virgo cluster, \cite{Cetal:94}
find two groups of galaxies with different \HI\ properties.  One group
has a typical \HI\ surface density ($\sim 4$ $M_{\sun}$ pc$^{-2}$) in
the inner regions, but a steeply falling one beyond a half of the
optical radius.  These galaxies could be affected by ram pressure
stripping.  The other group has lower central surface densities but
the gas extends much further, often beyond the optical radius.  In
other words, the latter galaxies are modestly deficient of \HI\ at all
radii, independently of the local gravitational force.  Other
evaporation mechanisms may play a role here, such as the turbulent
mixing or heat conduction (see \citealt{S:88} for details).

Also, in a study of three rich Abell clusters and the Virgo cluster,
\cite{VJ:91} find that the \HI\ deficiency is stronger in larger
spirals.  This statistical trend is opposite to that expected for ram
pressure stripping, since more massive galaxies should hold on to
their gaseous component better.  Having considered several possible
mechanisms, \cite{VJ:91} conclude that the \HI\ deficiency can only be
explained by tidal interactions.

Finally, note that numerical simulations of galaxy interactions
\citep{HM:95,BH:96} often show a dramatic loss of angular momentum of
the gas, drawing it to the center and igniting a powerful starburst.
This is corroborated by the observation that galaxies with distorted
profiles are more likely to be detected in H$\alpha$ than the
undistorted ones \citep{K:98}.  \cite{MW:93} find a 50\% higher
detection rate of Sa galaxies in the eight nearby Abell clusters than
in the field.  These starbursts may also blow away the remaining gas
from small galaxies \citep{DS:86}, leaving behind only compact stellar
spheroids.

\section{The Disruption of Low Density Galaxies}
  \label{sec:lsb}

\subsection{Dwarf Spheroidal Galaxies}

Dwarf galaxies are abundant in the Universe and dominate the number
counts both in the field and in clusters \citep{FB:94,GW:94}.  While
in the field they are usually found around large luminous galaxies,
their distribution in clusters is dramatically different.  There are
many fewer dwarfs in the center than on the periphery \citep{PDCS:98}
and there is a strong anti-correlation of the ratio of dwarfs to
giants with the local projected density of galaxies \citep{DCP:98}.
Also, less massive clusters are likely to have more dwarf galaxies
than large virialized clusters.  Observational evidence suggests that
dwarfs are subject to strong selection effects in clusters, consistent
with tidal disruption.

Dwarf galaxies are usually divided into two distinct classes, the high
density dE and the low surface brightness dSph \citep{K:85}.  It is
the dark matter-dominated dwarf spheroidals that are likely to be
disrupted.  For the purpose of this simulation I assume that initially
they were disk galaxies embedded in dark matter halos.  The surface
brightness profiles of many dSph are fairly good exponentials, which
supports their disk origin.  The typical scale length is about 1 kpc
\citep{FB:94}, similar to dwarf irregulars like the LMC
\citep{deVF:72}.  The velocity dispersion of dSph is very low,
typically 10 km s$^{-1}$ and no more than 20 km s$^{-1}$ \citep{K:85}.
I assume the latter extreme value to investigate whether even the
densest dwarf spheroidals could survive tidal heating.

I run three dwarf models \{D1, D2, D3\} with the tidal histories from
different cluster simulations (see \S\ref{sec:sample}).  Table
\ref{tab:initgal} shows that the stellar disks are small and are
completely dominated by dark matter.  Still, the density of the halo
cores is low, $\rho_{h,\rm av} \approx 5\times 10^{-3}\, M_{\sun}$
pc$^{-3}$, dangerously close to the critical tidal density
(eq. [\ref{eq:rhotid}]).  The peaks of the tidal force can reach as
high as 200 Gyr$^{-2}$, which corresponds to the tidal density of
$3.6\times 10^{-3}\, M_{\sun}$ pc$^{-3}$.

None of the dwarfs survives until the end of the simulation.  Galaxies
\{D1, D2, D3\} lose all of their halos and most of the stars in 6,
4.5, and 5 Gyr, respectively.  As is the case with the large spirals,
galaxy D2 from Cluster II experiences the strongest tidal heating,
while galaxy D1 from Cluster I the weakest.  The final state of the
models is very peculiar.  The particles are squeezed into a narrow
``tube'' stretching for several hundred kiloparsecs.  The shape of the
structure is truly one-dimensional; the axis ratios are smaller than a
few percent.  The orientation of the stream must be in the direction
of the strongest component of the tidal force at the moment of
disruption.  Of course, such a long straight line is an artifact of
the tidal approximation: in real clusters the tidal stream would bend
and curve beyond a few tens of kpc.

These results are not inconsistent with the presence of dwarf galaxies
in clusters.  A continuous infall of small galaxies is expected.
However once in the cluster, the diffuse dwarf spheroidals should be
disrupted in several crossing times.  In contrast, the dense dwarf
ellipticals should be almost unaffected by tidal heating and might be
found in any parts of the cluster.

\subsection{Low Surface Brightness Galaxies}

A much more luminous but even more diffuse type is the low surface
brightness galaxies.  LSB galaxies have slowly rising rotation curves
and very extended disks compared to the high surface brightness
spirals.  \cite{MdB:98} find $R_0 \sim 10$ kpc as a representative
scale length of galaxies in their sample.  Nevertheless, the
asymptotic values of the rotation curve of the high and low surface
brightness galaxies are similar, and they both follow the same
Tully-Fisher relation \citep{DSS:97,Detal:97}.

Similarly to dwarf spheroidals, the LSB galaxies are extremely
sensitive to tidal heating.  In order to investigate if they are also
prone to destruction, I run three models \{LSB1, LSB2, LSB3\} varying
the disk mass and scale length (Table \ref{tab:initgal}).  Each model
is embedded in a massive halo.  In model LSB1 the disk and the halo
have the same masses as in galaxy L3 but spread out with $R_0 = 10$
kpc.  In model LSB2 the disk mass is reduced by a factor of 4 relative
to LSB1.  And in the extreme model LSB3 the scale length is further
increased to $R_0 = 15$ kpc.  The tidal history of all LSB models is
the same as in galaxy L3.

Figure \ref{fig:vc_lsb0} shows the initial rotation curves for the
three LSB models.  Unlike Fig. \ref{fig:vc_ini}, the x-axis scale is
in kpc to emphasize the huge extent of these galaxies.  Even though
the halo mass is the same in all three models, the maximum value of
the circular velocity declines from 229 km s$^{-1}$ to 181 km s$^{-1}$
as a result of the decreasing disk mass and increasing scale length.
Dark matter dominates everywhere and determines the stability of
stellar disks against tidal disruption.  For galaxies LSB1 and LSB2
the average halo density in the core, $6 \times 10^{-3}\, M_{\sun}$
pc$^{-3}$, is above the critical tidal density, but galaxy LSB3 with
$\rho_{\rm av}(R_c) = 1.7 \times 10^{-3}\, M_{\sun}$ pc$^{-3}$ is not
expected to survive.

Simulations again confirm the predictions based on the tidal density.
Figure \ref{fig:denh_lsb} shows that the halos of models LSB1 and LSB2
are truncated at about 30 kpc, similarly to model L3.  For the LSB
galaxies, however, this implies losing 70\% of their dark mass (see
Table \ref{tab:fingal}).  As a result, the circular velocity falls by
18\% to 25\%.  Stellar disks are transformed even more strongly.
Although only 20\% of the stars are lost, the rest settle into a
spheroidal configuration with comparable axis ratios in all
directions.  The radial scale length shrinks, while the vertical scale
height expands by more than a factor of five.  The density
distribution is no longer exponential and declines more sharply
(Figure \ref{fig:dend_lsb}).  Any gas remaining in the diffuse disks
would be stripped by ram pressure of the intracluster media.  The
likely outcome of such evolution is a gas-devoid spheroidal galaxy.

The third model, LSB3, is essentially disrupted by tidal forces.  Only
4\% of the halo remains bound at the end of the simulation.  Because
of the larger initial extent and lower density, the truncation radius
is correspondingly smaller: $R_t \approx 30/1.5 = 20$ kpc (cf. Figure
\ref{fig:denh_lsb}).  The central density of the halo is reduced by
more than a factor of four.  Figure \ref{fig:dend_lsb} shows that the
stellar component is also severely damaged.  The disk shrinks to less
than a half of its original size and loses most of the stars.  The
surface density has fallen at all radii.  This small remnant of the
galaxy is unlikely to survive tidal heating for much longer and will
be torn apart by future tidal shocks.

To aid the comparison with observations, the critical tidal density
(eq. [\ref{eq:rhotid}]) can be translated to the projected surface
brightness of the disk that would be disrupted by tidal forces.  The
face-on disks with the scale length $R_0 \sim 3$ kpc would satisfy the
disruption condition $\rho_{\rm av}(R_0) < \rho_{\rm tid}$ if their
central surface brightness in the $B$ band is below
\begin{equation}
  \mu_{\rm tid} = 25.0 - 2.5 \log{
    \left({F_{\rm tid} \over 100\ \mbox{Gyr}^{-2}}\right)
    \left({R_0 \over 3 \ \mbox{kpc}}\right)
    \left({M/L_B \over 2}\right)^{-1}} \ \mbox{mag arcsec}^{-2},
\end{equation}
or $\mu_{\rm tid} = 23.8$ mag arcsec$^{-2}$ in the $R$ band.  For the
edge-on disks the critical surface brightness can be up to a factor of
10 higher.  However, if the dark matter cores are even denser
($\rho_{h,\rm av} > \rho_{d,\rm av}$) then $\mu_{\rm tid}$ would
correspondingly decrease.  The discoveries in the Coma cluster of the
giant diffuse arc with the surface brightness $\mu_B < 26.5$ mag
arcsec$^{-2}$ by \cite{TM:98} and the diffuse plume-like feature with
$\mu_R \approx 26$ mag arcsec$^{-2}$ by \cite{GW:98} may provide the
examples of a complete tidal disruption.

\subsection{Origin of Intracluster Light}
  \label{sec:intra}

Diffuse emission, not associated with galaxies, contributes 10\% to
50\% of the light from clusters \citep{Betal:95,VG:99}.  As the tidal forces
remove a substantial amount of stars from the low surface brightness
galaxies, tidal streams stretch across the clusters.  Is there enough
LSB galaxies to provide the origin of diffuse intracluster light?

The fraction of stars stripped from the galaxies can be estimated as
follows.  Assume that the initial galaxy luminosity function is given
by equation (\ref{eq:lumfun}) and that the luminosity, along with the
mass, scales linearly with the initial extent of the galaxy: $L
\propto R$.  Tidal truncation removes stars beyond the radius $R_t
\approx R_{t,*} \, (\sigma_g/\sigma_{g,*})$, where $R_{t,*}$ is the
tidal radius of $L_*$ galaxies.  Then the luminosity after truncation
is $L_t = L \, (R_t/R) = L_* \, (R_{t,*}/R_*) \,
(\sigma_g/\sigma_{g,*})$, and the fraction of dispersed light as a
function of $x \equiv L/L_*$ is
\begin{equation}
  f_d \equiv {1 \over L_{\rm tot}}
    {d L_{\rm lost} \over d \log{L}} 
    = {L - L_t \over L_{\rm tot}} \; N(L) L,
\end{equation}
where the total luminosity of the cluster is $L_{\rm tot} = \int N(L)
\, L \, dL = N_0 \, L_* \, \Gamma(2-\alpha)$.  Using the adopted
luminosity function and the Faber-Jackson relation $\sigma_g =
\sigma_{g,*} (L/L_*)^{1/4}$, the distribution becomes
\begin{equation}
  f_d(x) = {x^{1-\alpha} \over \Gamma(2-\alpha)}
           \left( x - {R_{t,*} \over R_{*}} x^{1/4} \right) \, e^{-x}.
\end{equation}
The simulations show that $R_{t,*} \approx 30$ kpc and the typical
ratio $R_{t,*}/R_{*} = 0.2 - 0.5$.  For $\alpha = 1 - 1.5$, the
distribution peaks around the characteristic scale $x=1$.

This calculation shows that most of the stripped stars come from giant
$L_*$ galaxies.  Since the giants dominate the integrated light of
both high and low surface brightness galaxies, the amount of the
intracluster light in the tidal disruption scenario is proportional to
the average fraction of LSB galaxies.  The comparison of the APM
survey and the CfA Redshift Survey \citep{Setal:97} shows that LSB
galaxies contribute about 20\% of the total light in the field.
\cite{Detal:97} find that this fraction may even be higher.  Also, in
the Fornax cluster the LSB galaxies contain 20\% to 30\% of the total
light \citep{K:00}.  Thus the disruption of large low surface
brightness galaxies would naturally account for the intracluster
light.

\section{Comparison with Previous Work}
  \label{sec:moore}

The idea of transforming spirals into S0s by tidal heating in clusters
is not new.  From the study of galaxy collisions in cluster cores,
\cite{R:76} finds that the dark matter halos are severely truncated.
Simulations of the tidal heating of colliding galaxies \citep{FS:81}
suggest that the disks grow in scale height, up to $z_0 \approx 0.2\,
R_0$.  Despite the limitations of analytical cluster models and small
number of particles, these early studies agree qualitatively with the
more advanced computations.

A detailed investigation of galaxies in clusters has not been possible
until very recently.  \cite{Metal:96} and \cite{MLK:98} have conducted
the first high-resolution simulation of the tidal effects, using a
singular isothermal cluster model populated by a large number of
tidally-truncated galaxies.  Both the neighbor galaxies and the
cluster halo exerted tides on the galaxy simulated with higher
resolution.  Moore et al. predicted a strong evolution: the disks of
medium size spirals ($V_c = 160 \ \mbox{km s}^{-1}$) were destroyed
and transformed into dwarf spheroidals.  Two effects may have
contributed to this: (i) they used the cluster with a higher velocity
dispersion (like Coma) where the tidal forces are stronger than in the
clusters (like Virgo) simulated here, and (ii) the clusters growing
hierarchically had less time to exert tidal effects than a steady
massive cluster.

More recent simulations of \cite{Moore:99} have incorporated the
hierarchical cluster formation in the $\Omega_0 = 1$ cosmology and
indicate more moderate tidal evolution.  They find that the disks of
high surface brightness spirals survive within their dark matter halos
but expand in vertical direction by a factor 2 to 3.  Their new
results are in a much better agreement with our models L1--L3.

Due to a smaller number of particles in the disk ($2\times 10^4$) and
in the halo ($5\times 10^4$), \cite{Moore:99} simulations could suffer
from numerical relaxation more strongly than our results.  Estimating
$m_h = 4\times 10^6\, M_{\sun}$, $\rho_h(R_c) \approx 0.2\, M_{\sun}$
pc$^{-3}$, $v_{\rm enc,h} \approx 180$ km s$^{-1}$ for their isolated
HSB galaxy, I find the disk heating time $t_{\rm rel,h} \sim 1.5$ Gyr
(cf. eq. [\ref{eq:trelh}]).  However, their Fig. 5 shows that in 5 Gyr
the disk has expanded by only slightly more than a factor of 2, which
corresponds to a relaxation time of 4.5 Gyr.  Thus for some reason, in
Moore's simulation the numerical relaxation rate is three times
slower.  The only worrying sign is the steepening of the central
density profile of the LSB galaxy in their Fig. 7.  Since tidal
heating vanishes in the center it cannot be responsible for this
effect, which could instead be a manifestation of the artificial
``core collapse''.

\cite{Moore:99} have examined the survival of the disks as a function
of their mass, scale length, and the core radius of the surrounding
dark matter halo.  They find that compact disks are stable regardless
of the halo core radius and that extended disks are stabilized by
denser halos.  All this can be explained in terms of a single tidal
density parameter (cf. eq. [\ref{eq:rhotid}]).  The galaxies are
stable if their combined inner density, of stars and dark matter, is
above the tidal density.  The results of Table 1 of \cite{Moore:99}
indicate the truncation radius of about 20 kpc for small halo core
radii ($R_c = 1$ kpc) and about 10 kpc for large core radii ($R_c =
10$ kpc), all of which correspond to a tidal density $\rho_{\rm tid}
\approx 3.6\times 10^{-3}\, M_{\sun}$ pc$^{-3}$.  (The fraction of
stars lost is independent of the disk mass as long as the dark halo
dominates, $M_d \ll M_h$.)

Despite different cluster models and simulation methods, the critical
tidal densities in \cite{Moore:99} and in this paper agree within a
factor of two, and the predicted dynamical evolution of galaxies is
also similar.  Some of the difference may be due to the close
encounters of neighbor galaxies within 60 $h^{-1}$ kpc, which are not
resolved in my cluster simulations (Paper I estimates that these
encounters contribute 10\% to 50\% of the overall tidal heating,
$I_{\rm tid}$, eq. [\ref{eq:ipar}]).  Nevertheless, it is remarkable
that such a simple concept as the tidal density of the Roche lobe
provides a quantitatively accurate description of detailed
simulations.  It offers a direct way to compare the results of
different authors.

Tidal disruption of galaxies as the origin of intracluster light has
been envisioned by \cite{M:84}.  He argued that the most probable time
for disruption is during the cluster collapse.  \cite{Moore:99} also
confirm the disruption of LSB galaxies in their simulations.
\cite{Cetal:99} present an excellent example of a diffuse arc
structure in the Centaurus cluster and explain it with the debris of a
low surface brightness galaxy disrupted by the giant elliptical NGC
4709.  Their finding that the stripped stars produce diffuse features
like those observed goes along with the results of \S\ref{sec:lsb}.
On a scale of the Local Group, \cite{Mayer:01} find that the observed
dwarf spheroidal galaxies could be the remnants of tidally-disrupted
LSB progenitors.

It should also be mentioned that \cite{D:98} has combined the
self-consistent cluster evolution with the galactic simulation in a
single parallel tree-code run.  He aimed to study the formation of a
giant cD galaxy in the cluster center through a sequence of mergers of
smaller galaxies.  Unfortunately, his approach does not allow enough
resolution to follow the dynamics of galaxies other than the central
giant.

\section{Discussion}
  \label{sec:dis}

The main result of this work is a natural and quantitative explanation
of the transformation of spiral galaxies into S0s.  The simulations
show that even on the orbits with maximum expected tidal heating,
large disk galaxies still remain flattened and not elliptical.  The
evolution of elliptical galaxies must have proceeded prior to the
cluster formation.

The main advance is made with the supporting simulations of the
hierarchical cluster formation, reported in Paper I.  The galactic
orbits, the tidal field along them, and the timescales of dynamical
evolution are all representative of the three plausible cosmological
scenarios of structure formation.  The evolution is stronger in the
models ($\Omega_0 < 1$) where galaxies spend more time in the dense
cluster environment.

Unlike previous studies with analytical cluster models, this work
demonstrates that tidal heating is not limited to the cluster core
because large local substructures are present in all parts of the
forming clusters.  In fact, most of the observable effects (halo
truncation and disk heating) happen in the outer regions, as soon as
the galaxy enters the cluster.  Also, most of the evolution can be
predicted using a single parameter, the critical tidal density along
the orbit (eq. [\ref{eq:rhotid}]).

A testable prediction of the tidally-driven evolution is the
preferential abundance of S0 galaxies in the middle of large clusters
at the expense of the spirals.  Disk galaxies on the cluster periphery
should have distorted profiles as well as the occasional spots of
diminishing star formation.  The abundance of elliptical galaxies, on
the other hand, should be universal.  These predictions are confirmed
by observations \citep{JSC:00}.  A systematic study of galaxies at
various stages of their evolution in clusters by the Sloan Digital Sky
Survey \citep{SDSS} will further improve the observational picture.

New HST images of the cluster CL 1358+62 \citep{vDetal:98} show that
the S0s are systematically bluer in the outer regions and may have
formed stars more recently than the inner S0s.  Interestingly, the
brightest blue galaxies are bulge-dominated.  Even though S0s are
systematically less luminous than both ellipticals and spirals
\citep{vdB:98}, their bulges may appear more prominent after star
formation has ceased: the disks would dim by up to 2 mag, while the
bulges would brighten due to the reduced extinction.  These effects,
however, may not be sufficient to explain the large bulges of S0s from
the small bulges of Sa-Sc galaxies \citep{Kent:85}.  Therefore, like
the elliptical galaxies, most of the bulges had to accumulate at the
early stages of galaxy formation.

The most significant limitation of the present study is the absence of
gas dynamics.  Future work should include a self-consistent treatment
of the gas, both in the cluster and the galaxy simulations.  Parallel
Tree-SPH codes should be able to address this issue, although the
amount of computations is very large.  Also, high-resolution cluster
simulations should help to resolve close galactic encounters.  A
nested grid method is necessary here, since the dynamical range should
be extended by at least a factor of 10, from 60 $h^{-1}$ kpc to less
than 5 $h^{-1}$ kpc.  Adaptive mesh codes, such as AMR \citep{AMR},
could reach the desired level but the computational expense is again
not currently possible.  The interpenetrating galaxy encounters may be
responsible for the unexpected star formation episodes, such as in the
``E+A'' galaxies \citep{G:89,Z:99}.

\section{Conclusions}

I have used the high-resolution \N-body simulations to explore the
dynamical evolution of disk galaxies in clusters.  The evolution is
driven by the tidal forces of the hierarchically forming clusters,
extracted from the lower resolution simulations (Gnedin 2003) in three
cosmological scenarios.  I find that tidal effects are capable of
inducing strong dynamical and morphological transformations, which can
be parametrized by the critical tidal density along the galactic
orbit.  In the Virgo-type clusters this tidal density is $\rho_{\rm
tid} \sim 2\times 10^{-3}\, M_{\sun}$ pc$^{-3}$.

(i) Disks of large spiral galaxies thicken by a factor of two to
three, while the stellar velocity dispersion in outer regions rises to
100 km s$^{-1}$.  Toomre's $Q$ parameter increases to $\gtrsim 2$ and
even a large amount of cold gas is unable to make the disks
gravitationally unstable.  The gas in the outer regions can be
stripped by ram pressure of the intracluster gas.  However, the
stellar surface density distribution remains similar to the initial.
Devoid of star formation these galaxies transform into the S0s, in
agreement with the observations of low redshift clusters.

(ii) Dark matter halos are truncated at $30 \pm 6$ kpc, for the
initial circular velocity of 250 km s$^{-1}$.  In smaller halos the
truncation radius is proportionately smaller, $R_t \propto V_c$.  As a
result, the galaxies with extended halos may lose most of their mass
to the cluster.

(iii) Low density (dwarf spheroidal and low surface brightness)
galaxies entering the cluster may get completely disrupted by tidal
shocks.  The stripped material would stretch in long tidal streams
along the original trajectories.  According to the current faint
galaxy surveys, the unbound stars of giant LSB galaxies may provide
enough material for the diffuse intracluster light.

(iv) In all three cosmological models tidal heating is not limited to
the cluster core and starts to affect galaxies as soon as they enter
the cluster.  The variation of the evolution on different orbits
within the same cluster is as large as the variation due to the
different cluster models.  Nevertheless, the epoch of cluster
formation, the critical tidal densities, and the heating rates
indicate that the low $\Omega_0$ models produce stronger dynamical
evolution of galaxies and agree with observations better than the
$\Omega_0=1$ model.

\acknowledgements

I would like to thank my thesis advisor J. P. Ostriker for his support
and guidance, Martin Weinberg for discussions, Mike Blanton for his
excellent Points visualization program, Lars Hernquist for the
routines generating disk initial conditions and for the original
version of the SCF code, and Jeremy Goodman, David Spergel, Neta
Bahcall, Scott Tremaine, and George Lake for valuable comments.  I
acknowledge the support from NSF grant AST 94-24416.  This work was
submitted in partial fulfillment of the Ph.D. requirements at
Princeton University.



\begin{figure}
  \label{fig:disk1}
\end{figure}

\onecolumn

  \vspace*{0.5cm}
  \centerline{\parbox{8.2cm}{\epsfig{file=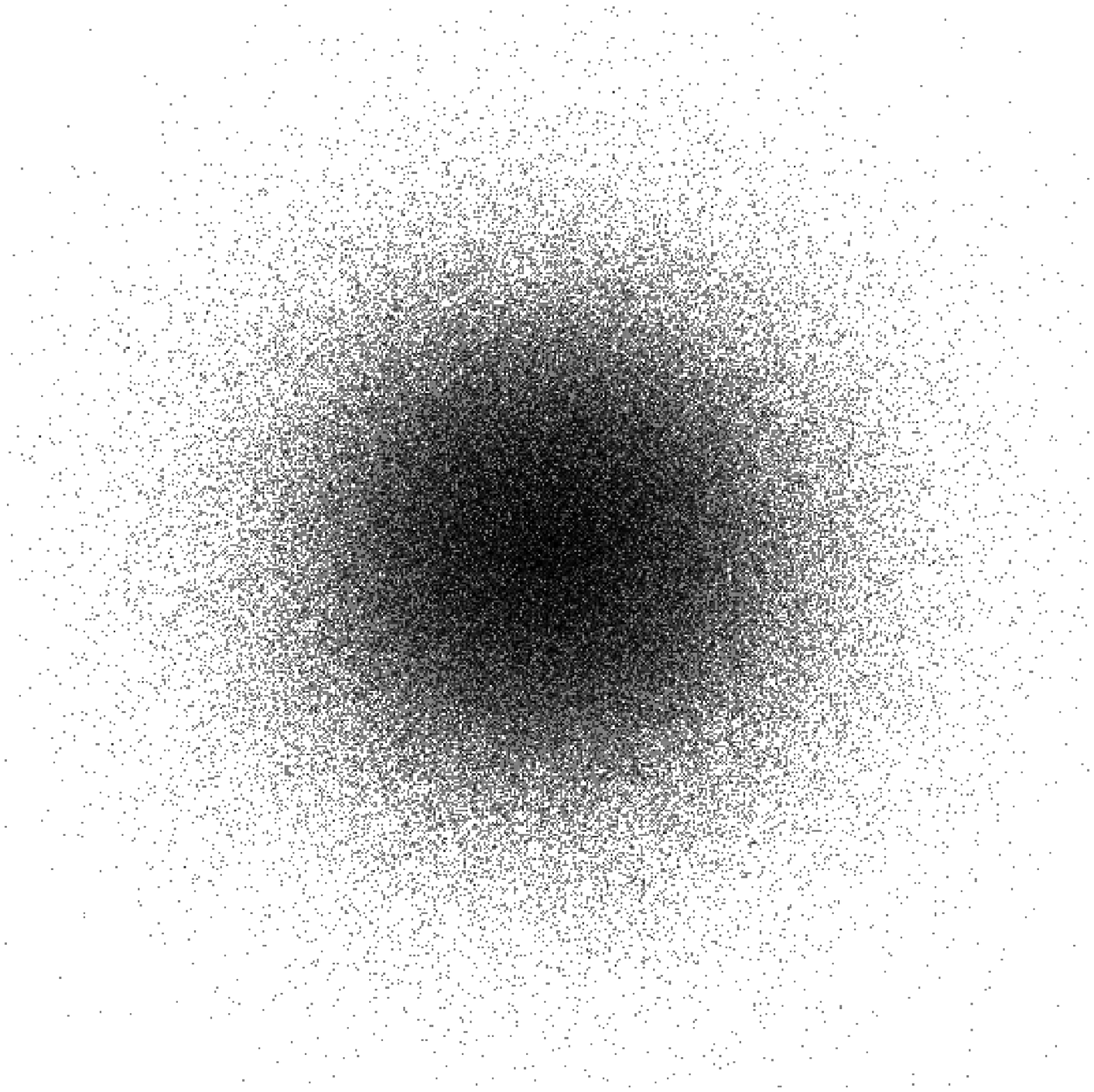,height=8.2cm}} \hspace{0.5cm}
              \parbox{8.2cm}{\epsfig{file=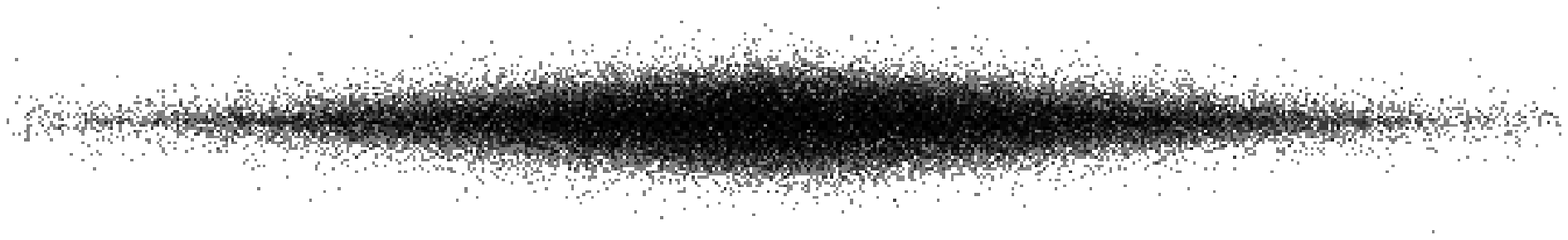,width=8.2cm}}}
  \vspace*{0.5cm}
  \centerline{\epsfig{file=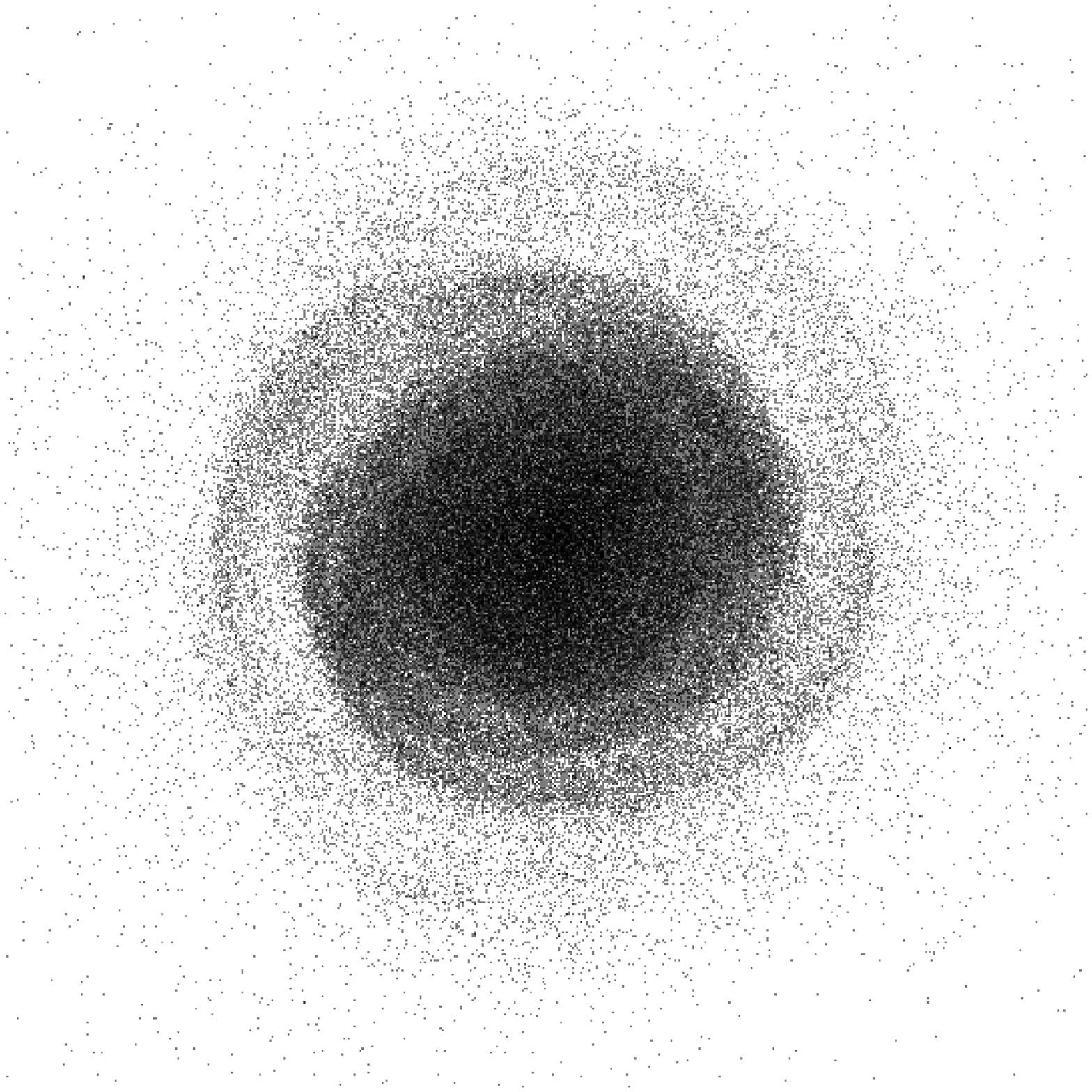,height=8.2cm} \hspace{0.5cm}
              \epsfig{file=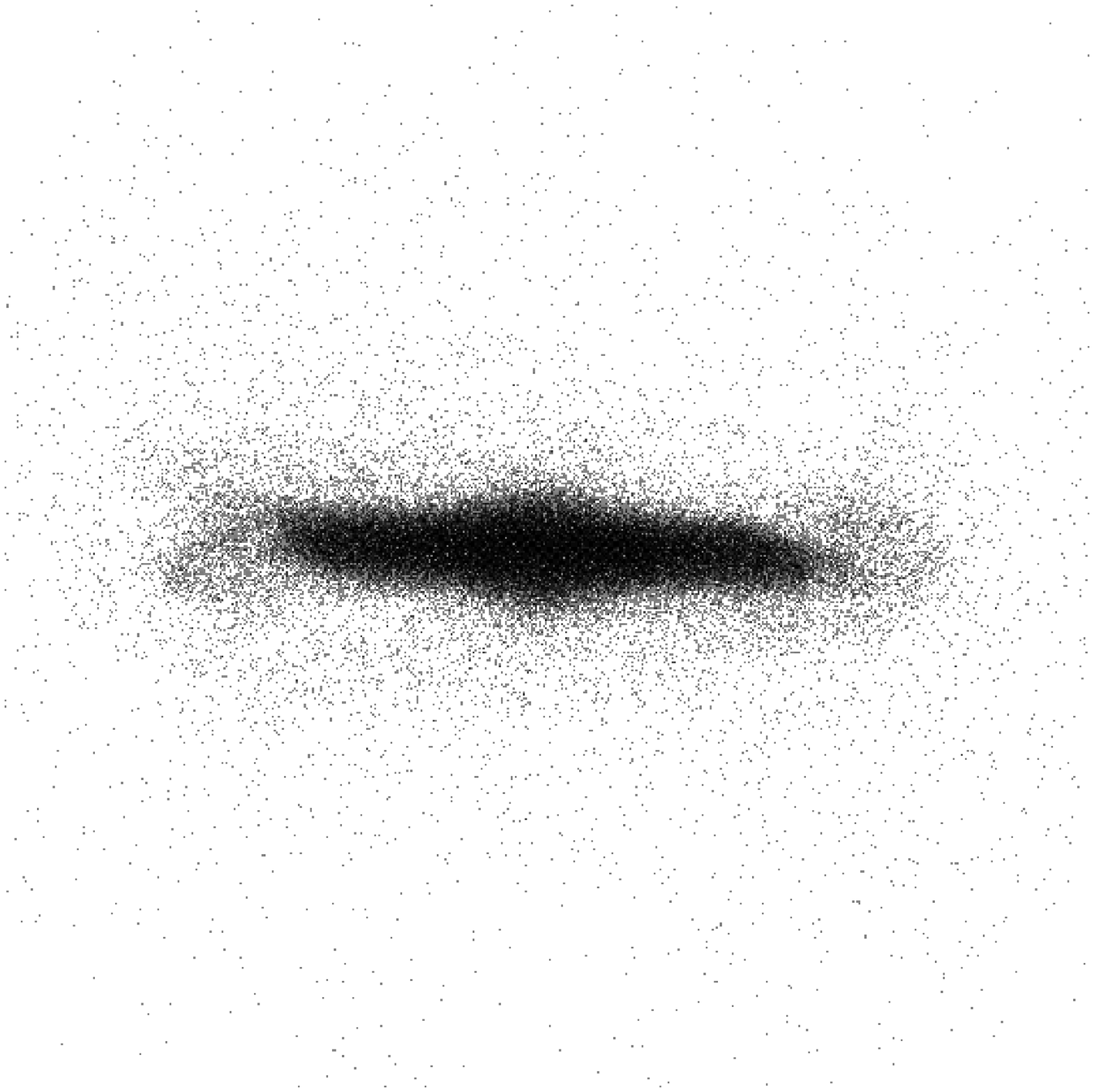,height=8.2cm}}
  \vspace*{1cm}
  \setcounter{figure}{2}
  \noindent 
  Fig. 2.---Upper panels: Initial conditions for large disk galaxies,
  face-on and edge-on views (only stellar particles are shown).
  Lower panels: Final configuration of galaxy L1.

\twocolumn

\begin{figure}
\plotone{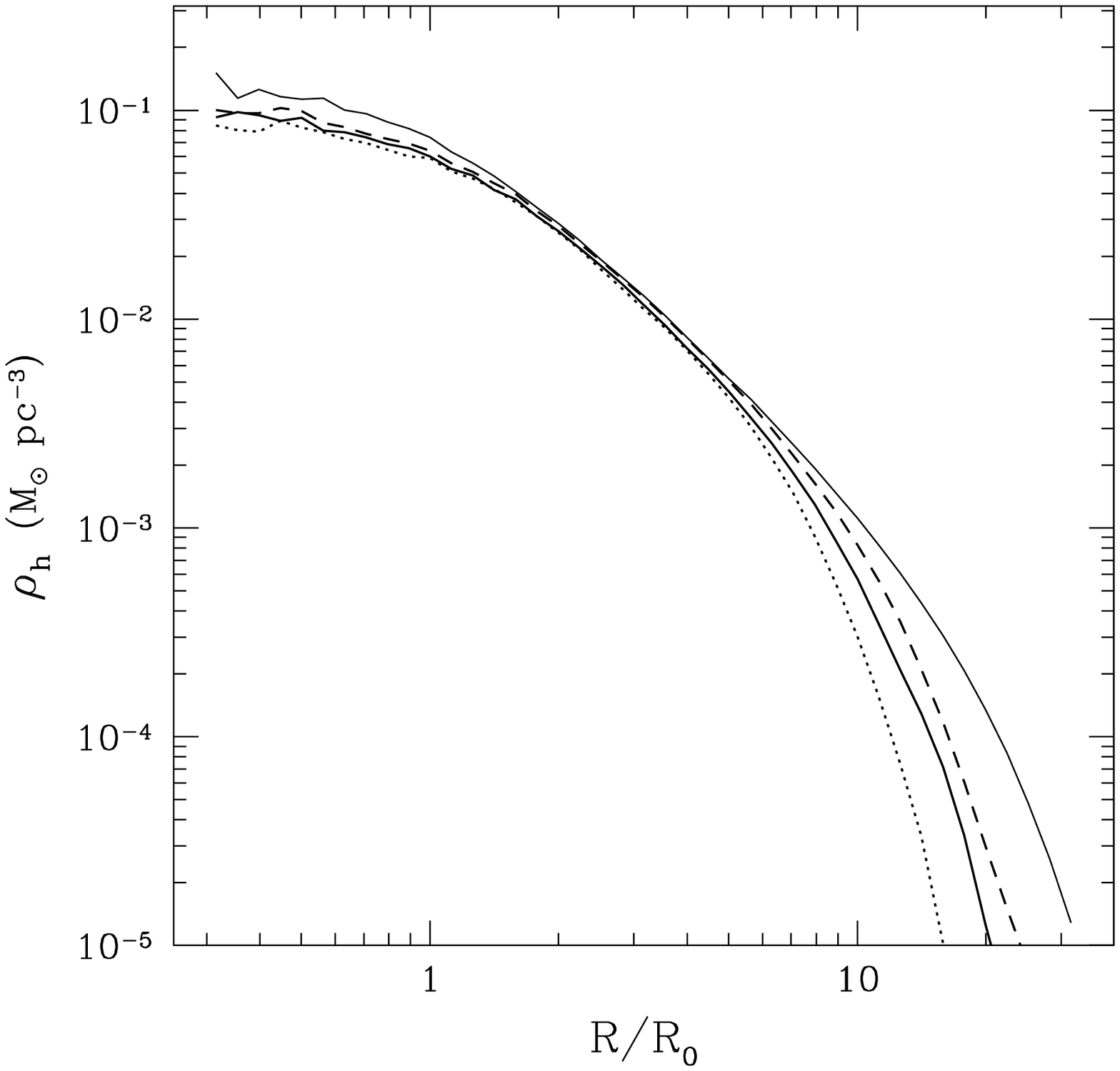}
\caption{Halo density profiles of large galaxies L1
  (dashes), L2 (dots), and L3 (solid line) at the end of each simulation.
  Outer thin line shows the initial distribution.
  \label{fig:denh}}
\end{figure}

\begin{figure}
\plotone{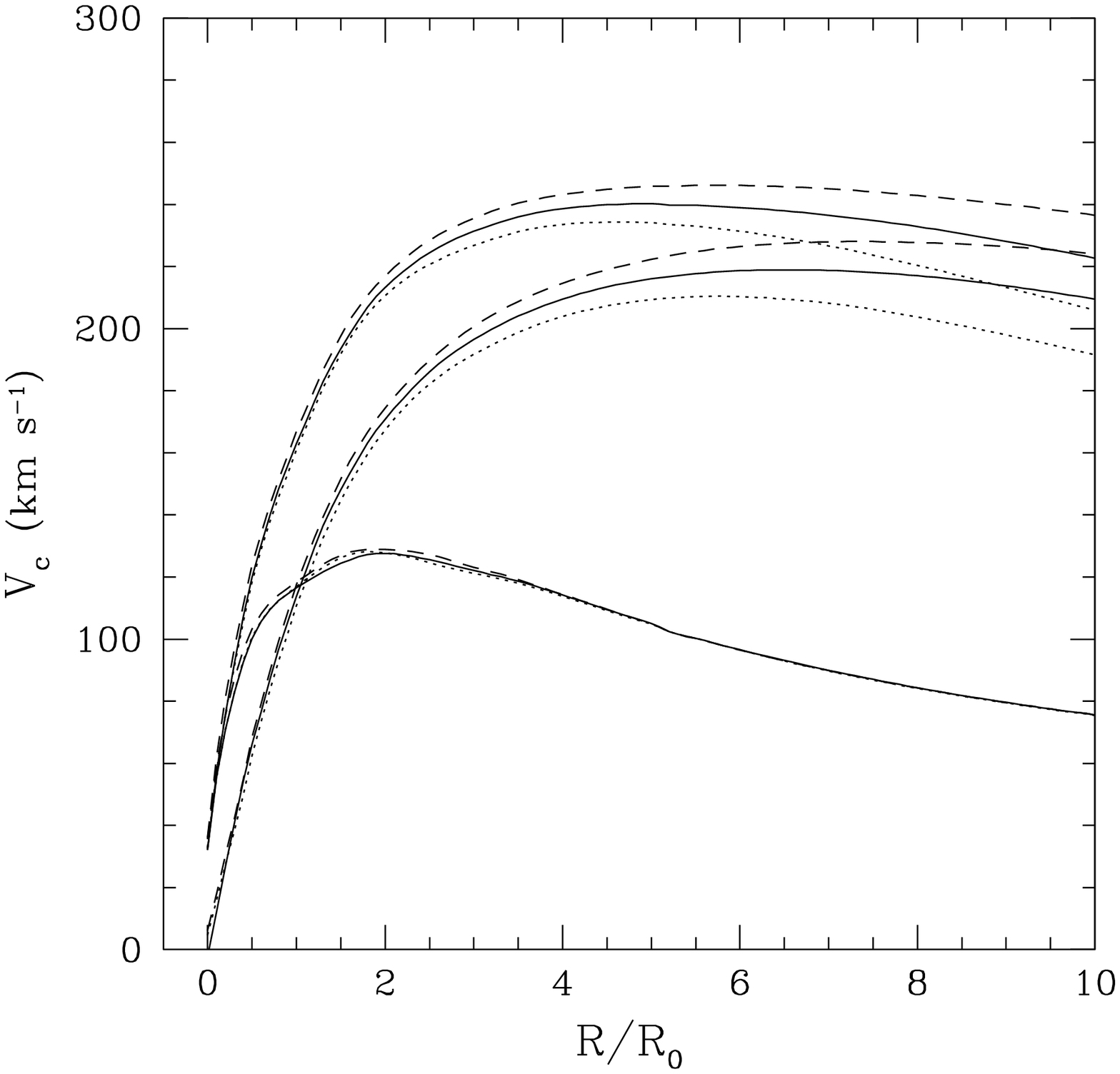}
\caption{Rotation curves, with the disk and halo
  contributions, of large galaxies L1 (dashes), L2 (dots), and L3
  (solid lines) at the end of each simulation.
  \label{fig:vcirc}}
\end{figure}

\begin{figure}
\plotone{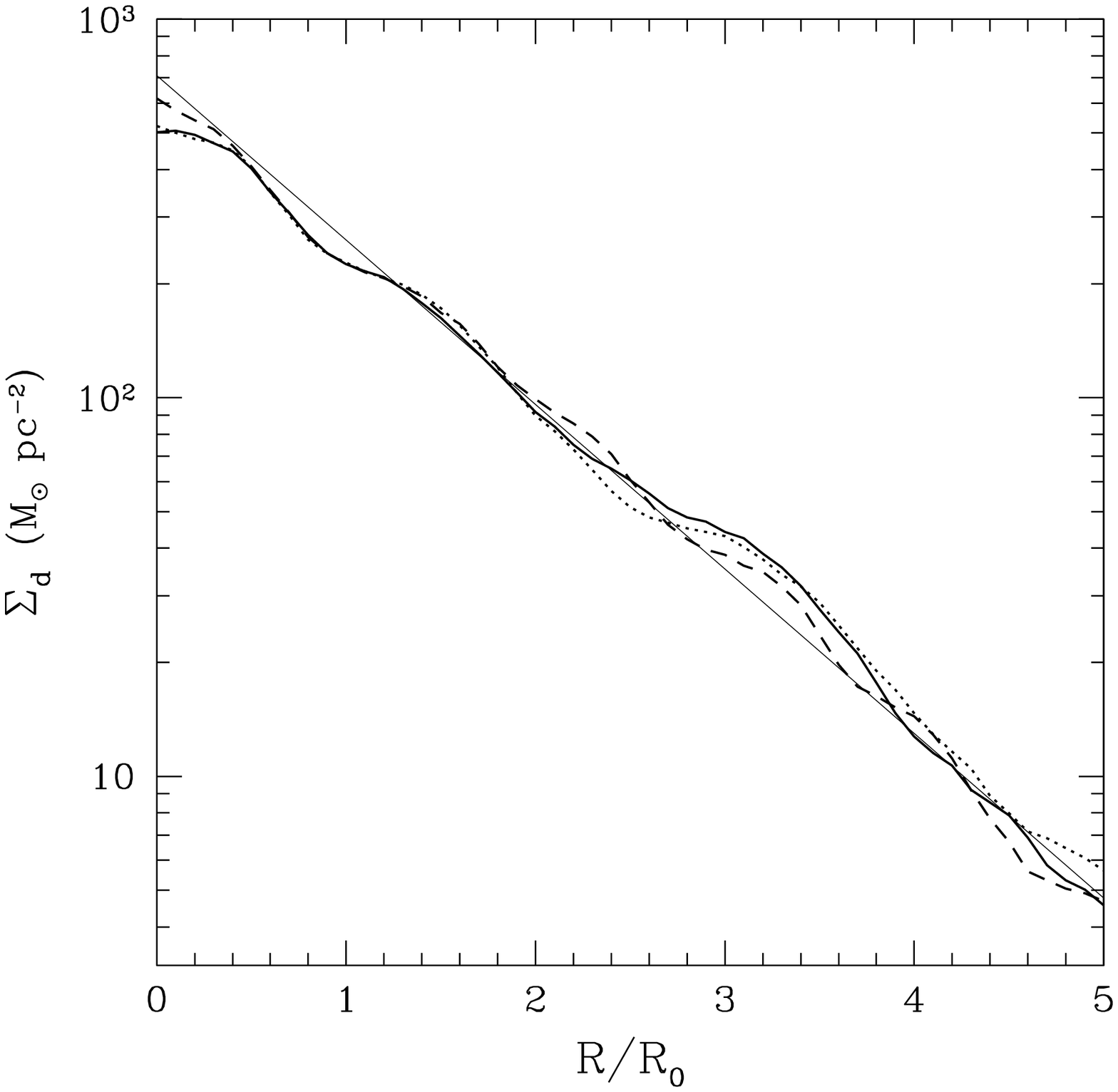}
\caption{Disk surface density of large galaxies L1 (dashes), L2
  (dots), and L3 (solid line) at the end of each simulation.  Straight
  thin line shows the initial exponential profile.
  \label{fig:dend}}
\end{figure}

\begin{figure}
\plotone{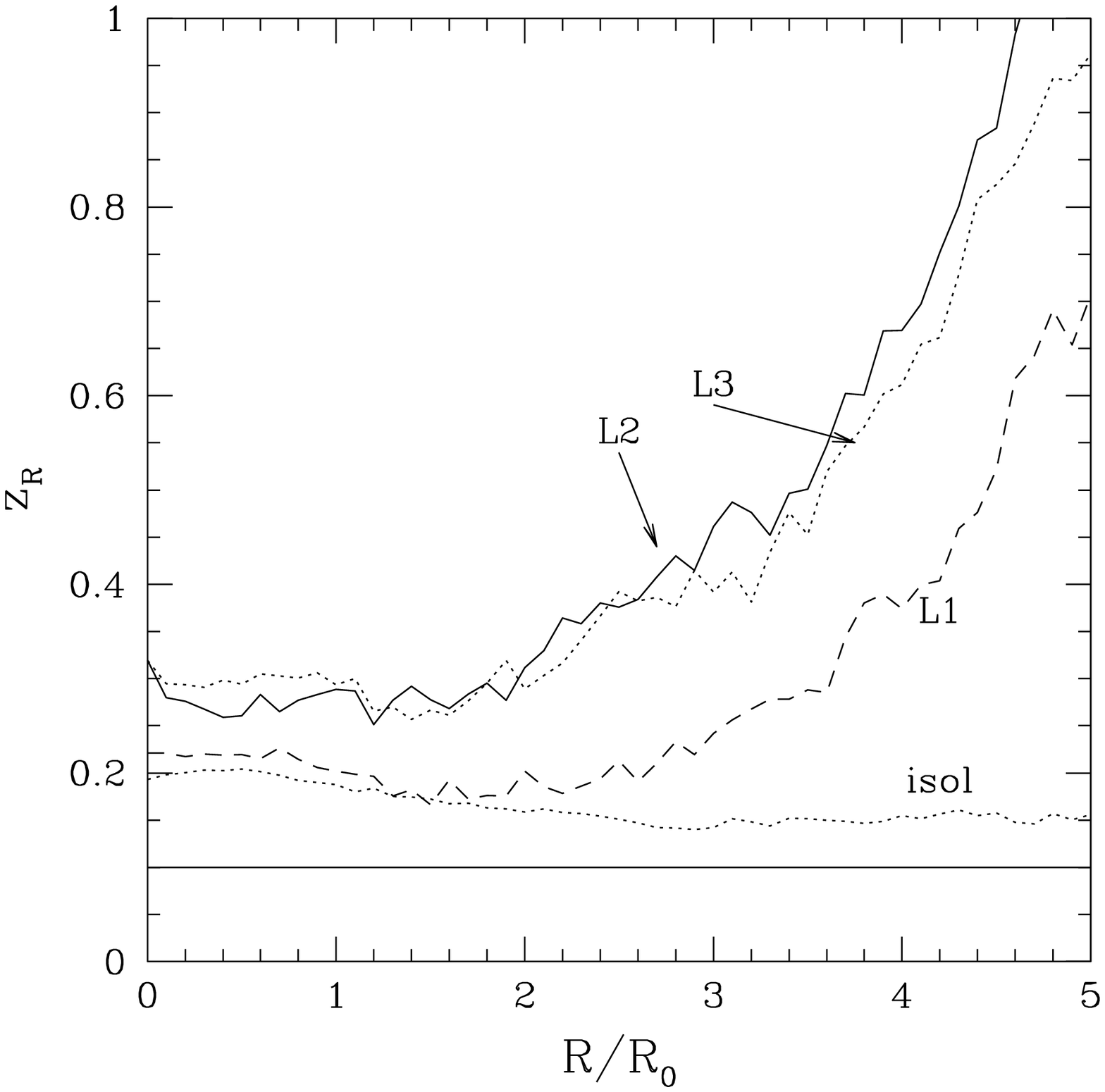}
\caption{Disk scale height, $z_0/R_0$, as a function of cylindrical radius
  for large galaxies L1 (dashes), L2 (dots), and L3 (solid line) at
  the end of each simulation.  Declining dotted line shows the
  numerical heating in the isolated run.  Solid horizontal line is
  the initial scale height, $0.1 \, R_0$.
  \label{fig:heatzr}}
\end{figure}

\begin{figure}
\plotone{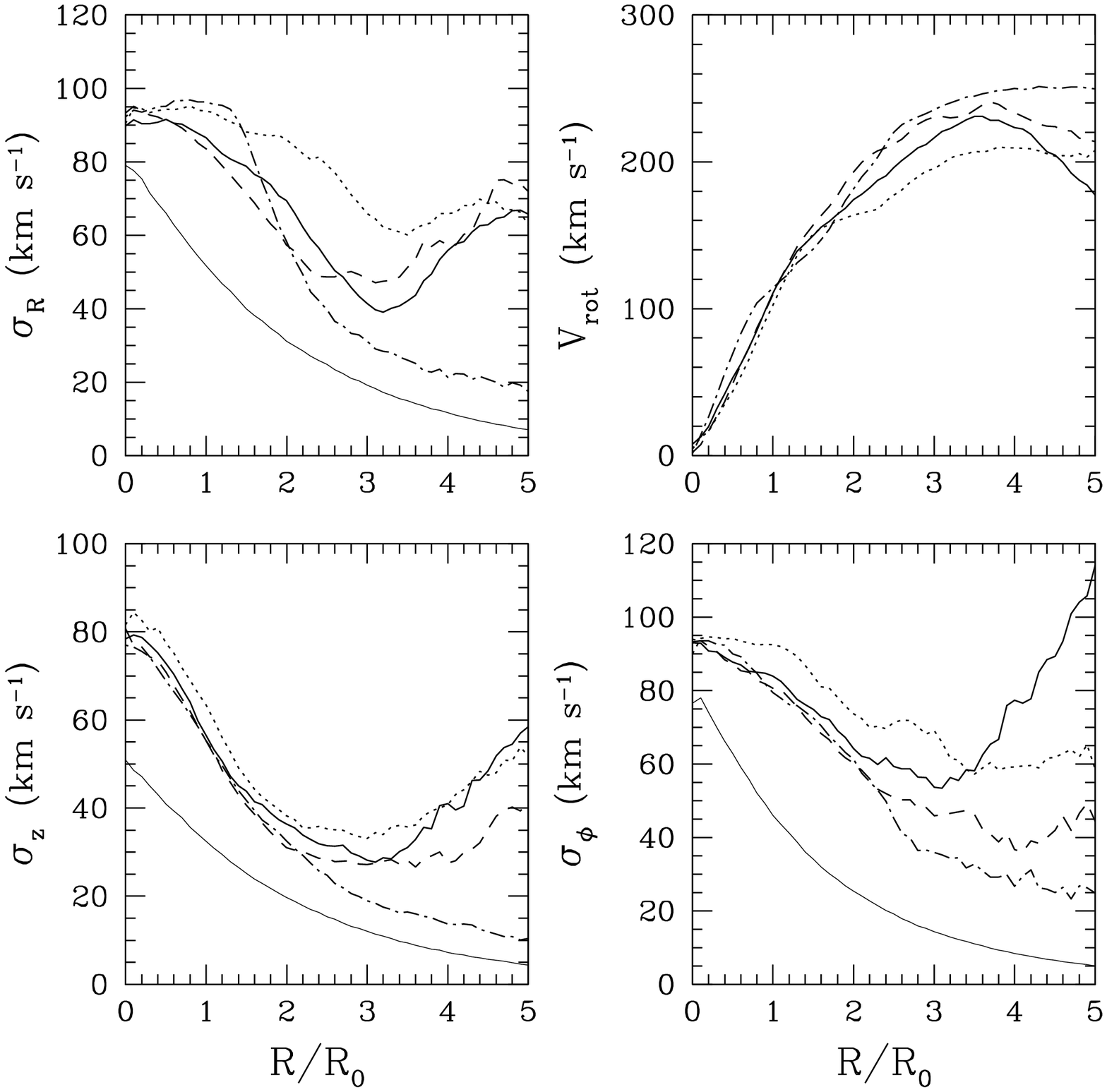}
\caption{Three components of the velocity dispersion, $\sigma_R$,
  $\sigma_\phi$, $\sigma_z$, and the rotation speed $V_{\rm rot}$ of the
  disks of large galaxies L1 (dashes), L2 (dots), and L3 (solid line) at
  the end of each simulation.  Dash-dotted line is for the isolated
  run.  Thin solid lines in the three dispersion panels show the
  initial state of the disk.
  \label{fig:dispd}}
\end{figure}

\begin{figure}
\plotone{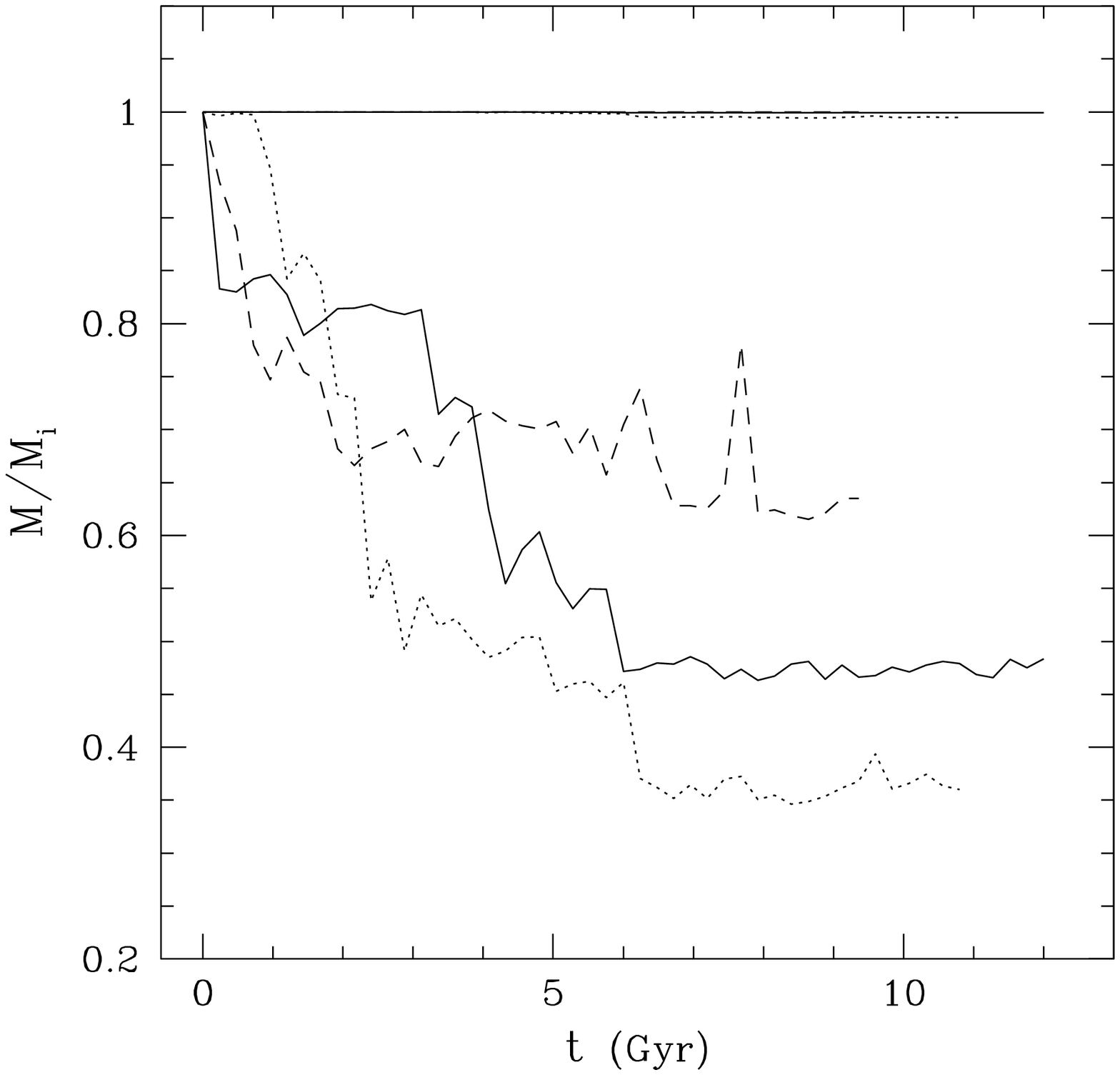}
\caption{The disk and halo masses of large galaxies L1 (dashes), L2
  (dots), and L3 (solid line) versus time.  Disk masses are essentially
  constant (horizontal lines).  Halo masses drop steadily (declining
  lines), although the instantaneous amount of bound mass may fluctuate
  with the external potential.  Note that simulation times
  differ for each cosmological model.  Sudden mass losses
  can be identified with the peaks of the external tidal force
  (cf. Figs. 14--16 in Paper I): at $t = 0.3, 0.8, 1.2$ Gyr for L1,
  at $t = 1.8, 2.2, 2.8, 6$ Gyr for L2,
  and at $t = 0.1, 1.4, 3.3, 4.2, 5.1$ Gyr for L3.
  \label{fig:halom}}
\end{figure}

\begin{figure}
\plotone{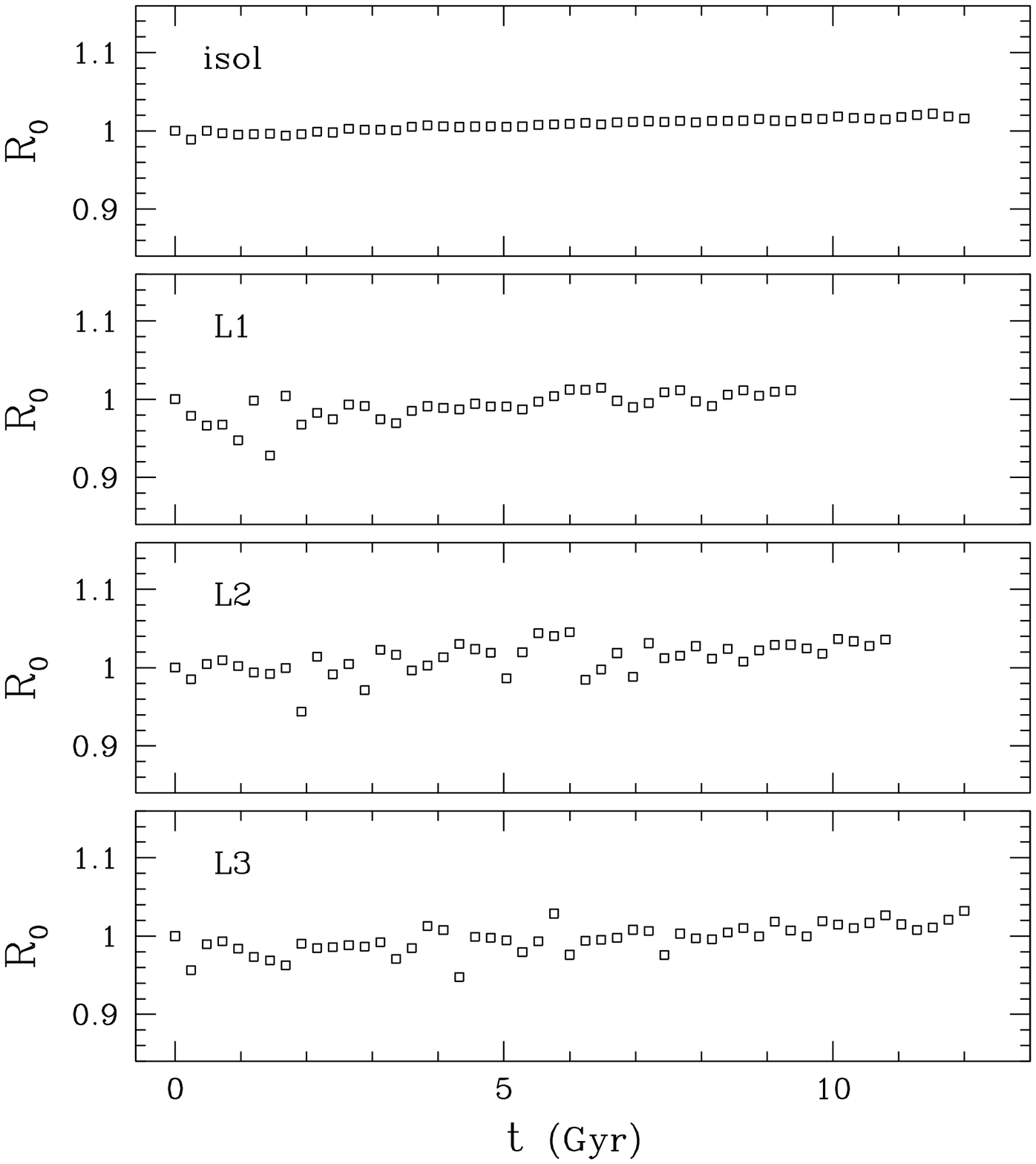}
\caption{Variation of the disk scale length with time for
  large galaxies L1, L2, L3 and for the isolated run.
  \label{fig:heatr}}
\end{figure}

\begin{figure}
\plotone{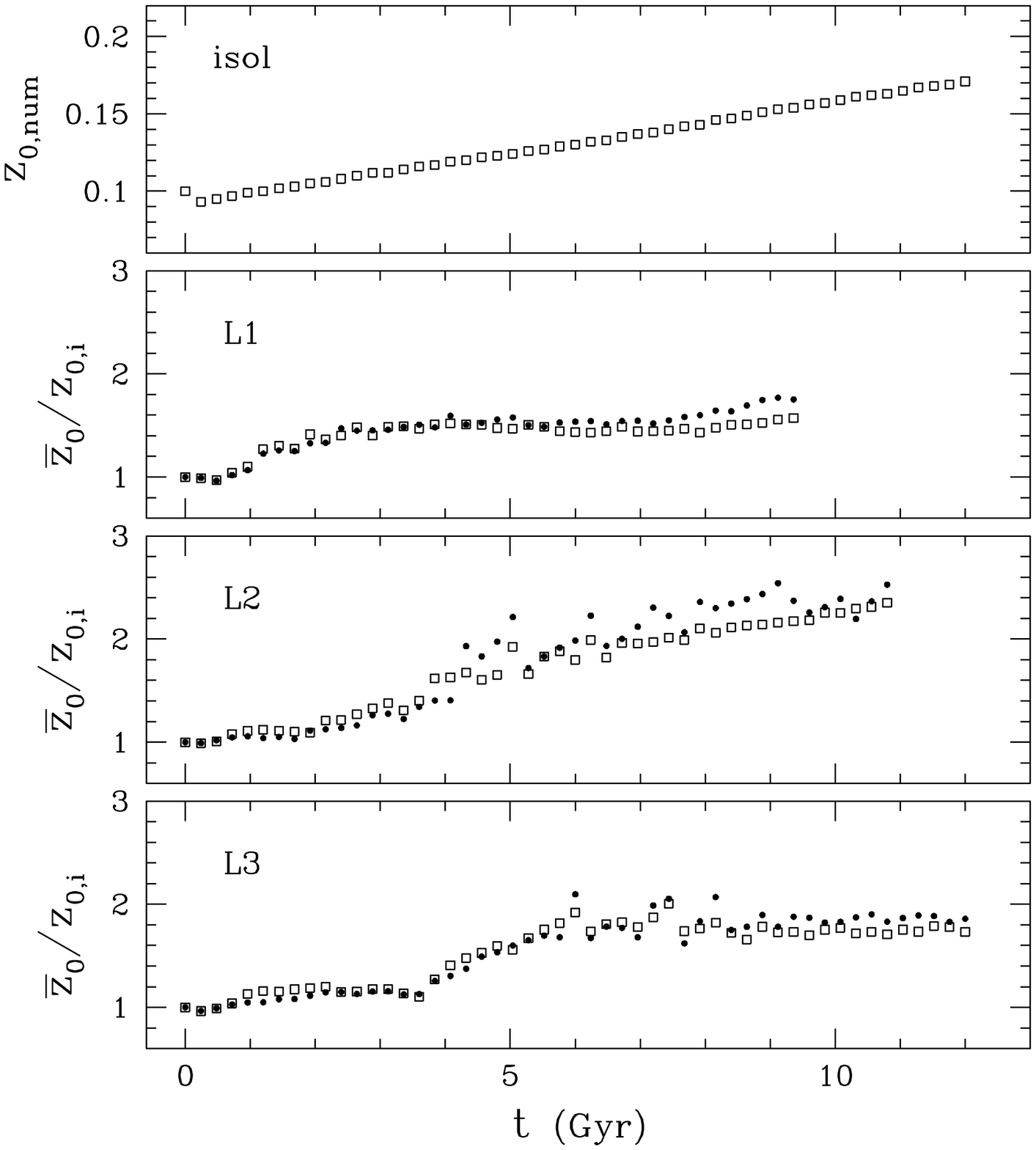}
\caption{Variation of the global scale height with time for
  large galaxies L1, L2, L3 and for the isolated run.  In the three
  lower panels, the instantaneous value of the scale height is normalized
  to the initial value and corrected for the effect of numerical heating
  (eq. [\protect\ref{eq:z0num}]).  Small dots show for comparison the
  results of \gadget\ simulations with more accurate vertical forces but
  with $N_d = 10^5$, $N_h = 4\times 10^5$ particles.
  \label{fig:heatz}}
\end{figure}

\begin{figure}
\plotone{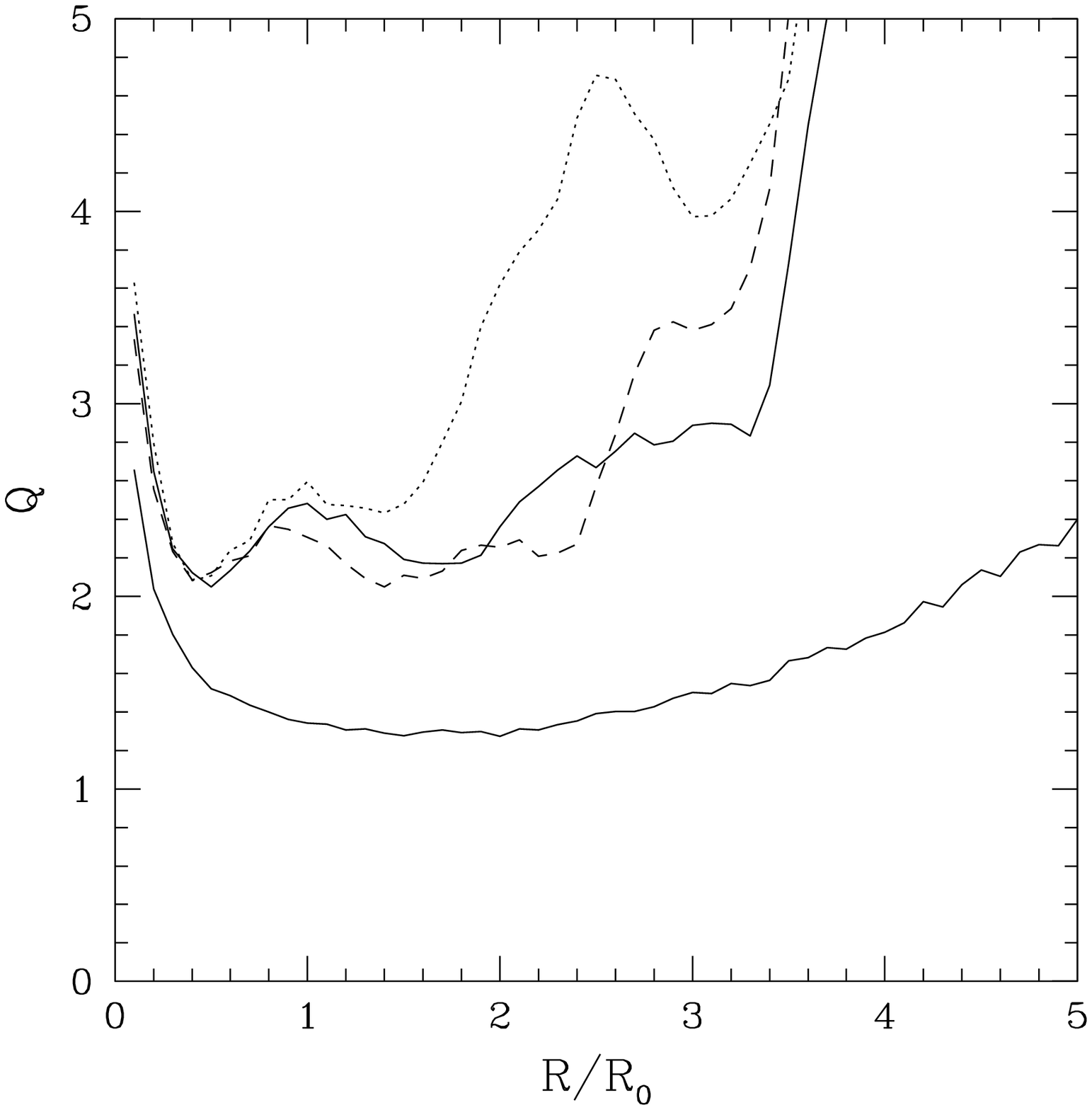}
\caption{Toomre's $Q$ parameter for the disks of large galaxies L1 (dashes),
  L2 (dots), L3 (solid line) at the end of each simulation.
  Lower solid line shows the initial state of the disk.
  \label{fig:q}}
\end{figure}

\begin{figure}
\plotone{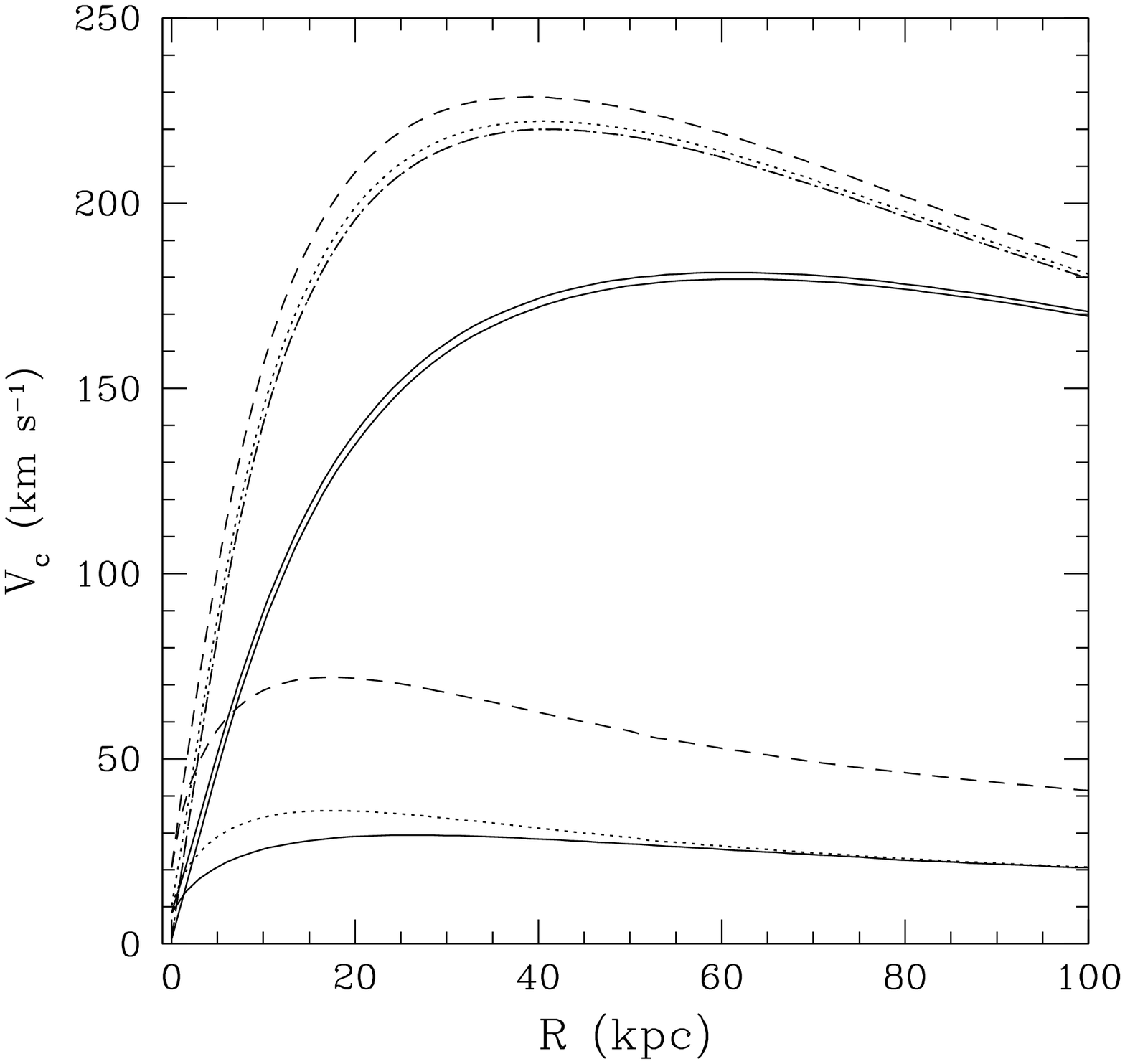}
\caption{Initial rotational curves, with the disk and halo
  contributions, of the low surface brightness galaxies
  LSB1 (dashes), LSB2 (dots), LSB3 (solid line).
  \label{fig:vc_lsb0}}
\end{figure}

\begin{figure}
\plotone{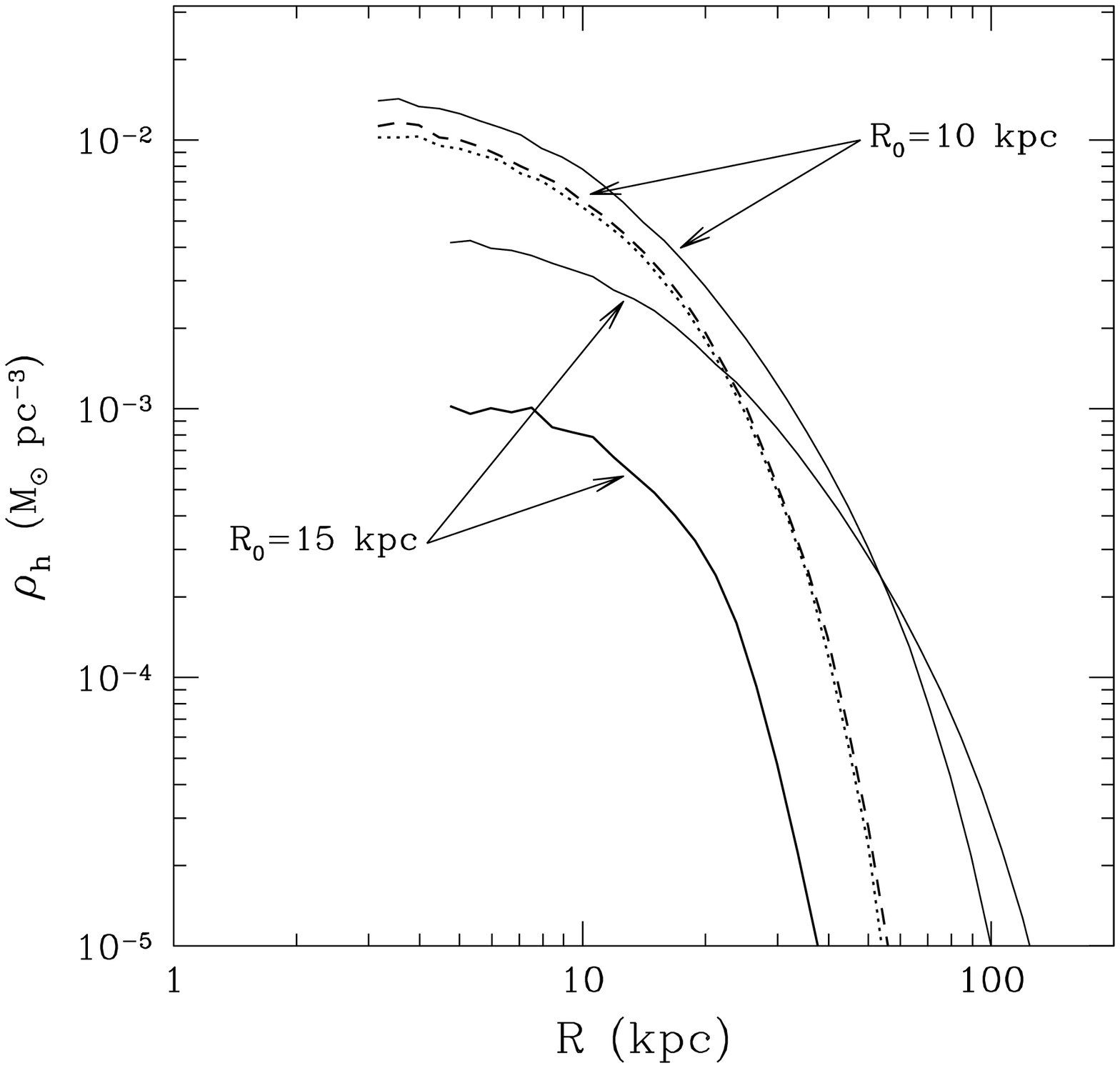}
\caption{Final halo density profiles of galaxies LSB1 (dashes),
  LSB2 (dots), LSB3 (thick solid line).  Thin solid lines show the
  initial profiles for each galaxy.
  \label{fig:denh_lsb}}
\end{figure}

\begin{figure}
\plotone{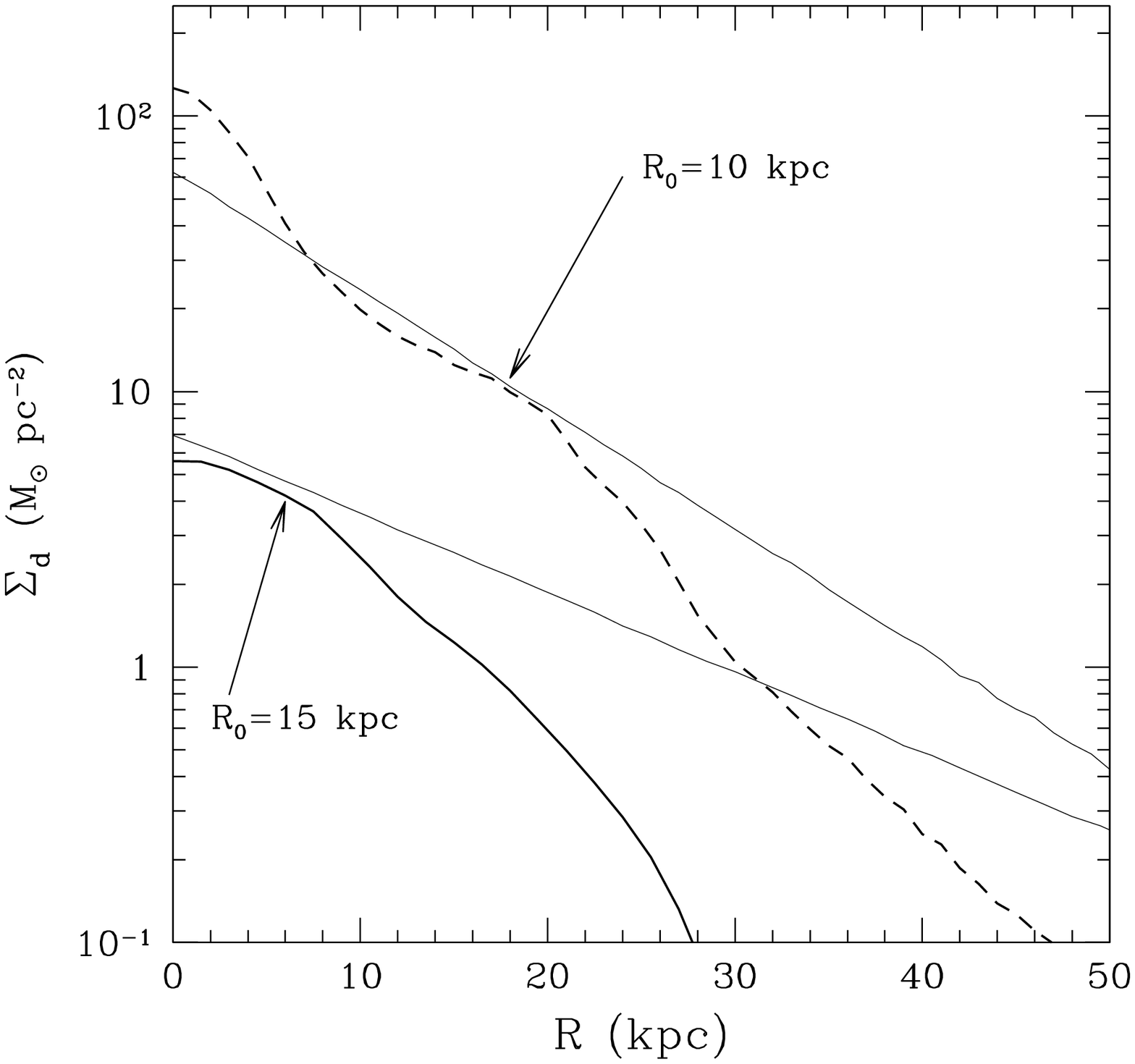}
\caption{Final disk surface density of galaxies LSB1 (dashes)
  and LSB3 (thick solid line).  Thin solid lines show the initial
  profiles for each galaxy.
  \label{fig:dend_lsb}}
\end{figure}


\clearpage

\begin{deluxetable}{lcccccc}
\tablecaption{Initial parameters of simulated galaxies \label{tab:initgal}}
\tablecolumns{7}
\tablehead{\colhead{} & \colhead{$M_h$} & \colhead{$M_d$} &
           \colhead{$V_c$} & \colhead{$R_0$} & \colhead{$z_0$} &
           \colhead{$T_{\rm rot}$}\\
           \colhead{Model} & \colhead{($M_{\sun}$)} & \colhead{($M_{\sun}$)} &
           \colhead{(km s$^{-1}$)} & \colhead{(kpc)} & \colhead{(kpc)} &
           \colhead{(yr)}}
\tablewidth{0pt}
\startdata
L1,L2,L3     & $8\times 10^{11}$ & $4\times 10^{10}$ & 250 & 3 & 0.3 &
  $1.2\times 10^8$ \\
D1,D2,D3     & $2\times 10^9$    & $10^8$            &  20 & 1 & 0.1 &
  $5.0\times 10^8$ \\
LSB1         & $8\times 10^{11}$ & $4\times 10^{10}$ & 229 & 10 & 1 &
  $4.6\times 10^8$ \\
LSB2         & $8\times 10^{11}$ & $10^{10}$         & 222 & 10 & 1 &
  $4.7\times 10^8$ \\
LSB3         & $8\times 10^{11}$ & $10^{10}$         & 181 & 15 & 1.5 &
  $7.0\times 10^8$
\enddata
\end{deluxetable}

\begin{deluxetable}{lccccccc}
\tablecaption{Isolated runs\label{tab:isola}}
\tablewidth{0pt}
\tablenum{2a}
\tablecolumns{8}
\tablehead{\colhead{Model} & \colhead{$N_d$} & \colhead{$N_h$} &
           \colhead{$\Delta t$ (yr)} &
           \colhead{$R_0/R_{0,i}$} & \colhead{$z_0/z_{0,i}$} &
           \colhead{$b/a$} & \colhead{$c/a$}}
\startdata
L--isol    & $10^6$ & $10^6$ & $1.2\times 10^6$ & 
             1.04 & 1.7 & 0.99 & 0.97 \\
L--isol1   & $10^6$ & $10^6$ & $2.4\times 10^6$ & 
             1.06 & 2.1 & 1.00 & 0.97 \\
L--isol2   & $5\times 10^5$ & $5\times 10^5$ & $1.2\times 10^6$ &
             1.04 & 1.9 & 1.00 & 0.97 \\
L--isol3   & $5\times 10^5$ & $10^6$ & $1.2\times 10^6$ &
             1.04 & 1.9 & 0.99 & 0.97 \\
LSB--isol3 & $10^6$ & $10^6$ & $7.0\times 10^6$ &
             1.01 & 0.8 & 1.00 & 0.88
\enddata
\end{deluxetable}

\begin{deluxetable}{lcccc}
\tablecaption{Energy changes in isolated runs\label{tab:isolb}}
\tablewidth{0pt}
\tablenum{2b}
\tablecolumns{5}
\tablehead{\colhead{Model} &
           \colhead{$\left<{\Delta E/E}\right>_d$} &
           \colhead{$\left<{(\Delta E/E)^2}\right>_d$} &
           \colhead{$\left<{(\Delta E/E)^2}\right>_h$} &
           \colhead{$\left<{\Delta E_z}\right>/\left<{E_z}\right>_d$}}
\startdata
L--isol    & $1.3\times 10^{-3}$  & $4.8\times 10^{-3}$ & $1.3\times 10^{-3}$
           & 1.0 \\
L--isol1   & $-1.5\times 10^{-2}$ & $1.1\times 10^{-2}$ & $2.1\times 10^{-3}$
           & 1.1 \\
L--isol2   & $2.6\times 10^{-3}$  & $7.2\times 10^{-3}$ & $1.7\times 10^{-3}$
           & 1.1 \\
L--isol3   & $4.9\times 10^{-3}$  & $7.9\times 10^{-3}$ & $1.6\times 10^{-3}$
           & 1.2 \\
LSB--isol3 & $3.2\times 10^{-2}$  & $3.1\times 10^{-4}$ & $1.2\times 10^{-3}$
           & 0.55
\enddata
\end{deluxetable}

\setcounter{table}{2}

\begin{deluxetable}{lcccccccc}
\tablecaption{Final parameters of large galaxies \label{tab:fingal}}
\tablewidth{0pt}
\tablecolumns{9}
\tablehead{\colhead{Model} & \colhead{$T_{\rm sim}$ (Gyr)} &
           \colhead{$M_h/M_{h,i}$} &
           \colhead{$M_d/M_{d,i}$} & \colhead{$V_c/V_{c,i}$} &
           \colhead{$R_0/R_{0,i}$} & \colhead{$\bar{z}_0/z_{0,i}$} &
           \colhead{$b/a$} & \colhead{$c/a$}}
\startdata
L1     &  9.35 & 0.63 & 1.000 & 0.98 & 1.01 & 1.6 & 0.93 & 0.88 \\
L2     & 10.71 & 0.36 & 0.995 & 0.94 & 1.04 & 2.4 & 0.93 & 0.88 \\
L3     & 12.28 & 0.48 & 0.999 & 0.95 & 1.03 & 1.7 & 0.89 & 0.84 \\
\hline
L3a    & 12.28 & 0.55 & 1.000 & 0.99 & 1.01 & 1.4 & 0.92 & 0.87 \\
L3b    & 12.28 & 0.39 & 0.997 & 0.93 & 1.02 & 2.5 & 0.91 & 0.88 \\
\hline
LSB1   & 12.28 & 0.34 & 0.81  & 0.82 & 0.74 & 5.7 & 0.86 & 0.84 \\
LSB2   & 12.28 & 0.32 & 0.77  & 0.75 & 0.78 & 5.2 & 0.87 & 0.79 \\
LSB3   & 12.28 & 0.04 & 0.28  & 0.29 & 0.43 & 2.4 & 0.79 & 0.72
\enddata
\end{deluxetable}

\begin{deluxetable}{lcccccccccc}
\small
\tablewidth{0pt}
\tablecolumns{11}
\tablecaption{Test models of galaxy L3\label{tab:L3gal}}
\tablehead{\colhead{Model} & \colhead{$N_d$} & \colhead{$N_h$} &
           \colhead{$\Delta t$ (yr)} &
           \colhead{$M_h/M_{h,i}$} &
           \colhead{$M_d/M_{d,i}$} & \colhead{$V_c/V_{c,i}$} &
           \colhead{$R_0/R_{0,i}$} & \colhead{$\bar{z}_0/z_{0,i}$} &
           \colhead{$b/a$} & \colhead{$c/a$}}
\startdata
main     & $10^6$ & $10^6$ & $1.2\times 10^6$ &
           0.48 & 0.999 & 0.95 & 1.03 & 1.7 & 0.89 & 0.84 \\
1        & $10^6$ & $10^6$ & $2.4\times 10^6$ & 
           0.49 & 0.998 & 0.95 & 1.08 & 1.9 & 0.89 & 0.85 \\
2        & $5\times 10^5$ & $5\times 10^5$ & $1.2\times 10^6$ &
           0.48 & 0.999 & 0.95 & 1.04 & 1.7 & 0.89 & 0.85 \\
3        & $5\times 10^5$ & $10^6$ & $1.2\times 10^6$ &
           0.48 & 0.999 & 0.95 & 1.04 & 1.5 & 0.89 & 0.84
\enddata
\end{deluxetable}

\begin{deluxetable}{lccccccccc}
\tablecaption{\gadget\ runs\label{tab:gadget}}
\tablewidth{0pt}
\tablecolumns{10}
\tablehead{\colhead{Model} & \colhead{$N_d$} & \colhead{$N_h$} &
           \colhead{$M_h/M_{h,i}$} &
           \colhead{$M_d/M_{d,i}$} & \colhead{$V_c/V_{c,i}$} &
           \colhead{$R_0/R_{0,i}$} & \colhead{$\bar{z}_0/z_{0,i}$} &
           \colhead{$b/a$} & \colhead{$c/a$}}
\startdata
L1-\gadget & $10^5$ & $10^5$ & 0.63 & 0.999 & 0.90 & 1.15 & 2.1 & 0.90 & 0.89 \\
L2-\gadget & $10^5$ & $10^5$ & 0.36 & 0.994 & 0.84 & 1.10 & 3.1 & 0.94 & 0.87 \\
L3-\gadget & $10^5$ & $10^5$ & 0.45 & 0.999 & 0.83 & 1.18 & 2.8 & 0.93 & 0.91 \\
\hline
L1-\gadget & $10^5$ & $4\times 10^5$ 
                             & 0.63 & 0.999 & 0.90 & 1.12 & 1.8 & 0.90 & 0.89 \\
L2-\gadget & $10^5$ & $4\times 10^5$ 
                             & 0.36 & 0.994 & 0.84 & 1.11 & 2.5 & 0.95 & 0.88 \\
L3-\gadget & $10^5$ & $4\times 10^5$ 
                             & 0.45 & 0.999 & 0.83 & 1.20 & 1.9 & 0.92 & 0.90
\enddata
\end{deluxetable}

\end{document}